\documentclass[twocolumn, preprintnumbers, amsmath, amssymb,nofootinbib]{revtex4}
\usepackage{tabularx, graphicx}
\usepackage{xfrac}
\usepackage{color}
\usepackage{hyperref}
\usepackage{soul}
\usepackage{ulem} 
\hypersetup{colorlinks=true, linkcolor=blue, filecolor=blue, urlcolor=blue,}

\usepackage{color}

\begin{document}
	\newcommand{\beq}{\begin{equation}}
		\newcommand{\eeq}{\end{equation}}
	\newcommand{\beqn}{\begin{eqnarray}}
		\newcommand{\eeqn}{\end{eqnarray}}
	\newcommand{\bmath}{\begin{subequations}}
		\newcommand{\emath}{\end{subequations}}
	\newcommand{\bra}[1]{\langle #1|}
	\newcommand{\ket}[1]{|#1\rangle}
	
	\title{Analysis of ``Revaluation of the lower critical field in
		superconducting $H_{3}S$ and $LaH_{10}$ (Nature Comm. 13, 3194, 2022)'' by V.
		S. Minkov et al}
	
	\author{J. E. Hirsch$^{a}$ and M. van Kampen$^{b}$ }
	\address{$^{a}$Department of Physics, University of California, San Diego, La Jolla, CA 92093-0319\\ $^{b}$ForBetterScience, Erlensee, Germany}
	
	\begin{abstract}
		In \href{https://www.nature.com/articles/s41467-022-30782-x}{Nat Comm. 13,
			3194, 2022} \cite{e2021p} and an ``\href{https://www.nature.com/articles/s41467-023-40837-2}{Author
			Correction}'' to it \cite{correction}, Minkov et al. presented magnetization
		data versus applied magnetic field for $H_{3}S$ and $LaH_{10}$ under pressure,
		argued that the data provide evidence that these materials are
		superconducting at high temperatures, and extracted from the reported data the
		behavior of lower critical fields versus temperature. In several papers
		\cite{unpulling,bending, hysteresis,ma} analyzing Refs. \cite{e2021p,correction}
		it was shown that the published magnetization data could not have been obtained
		from the reported measured data through the processes described in Refs.
		\cite{e2021p,correction}. Recently, Minkov et al performed a revaluation of
		their experimental results \cite{reval} and argued that the results derived
		from their new analysis are consistent with the results reported earlier
		\cite{e2021p}. In addition, they made public the underlying data \cite{data}
		from which the data published in Ref. \cite{e2021p} were derived. In this
		paper we analyze those underlying data and conclude that (a) the data
		published in Ref. \cite{e2021p} are incompatible with the underlying measured
		data, and (b) the revaluation analysis presented in Ref. \cite{reval} does not
		support the conclusions drawn by the authors in Ref. \cite{reval} nor Ref. \cite{e2021p}.
	\end{abstract}
	
	\maketitle
	
	\begin{figure}[t]
		\resizebox{8.2cm}{!}{\includegraphics[width=6cm]{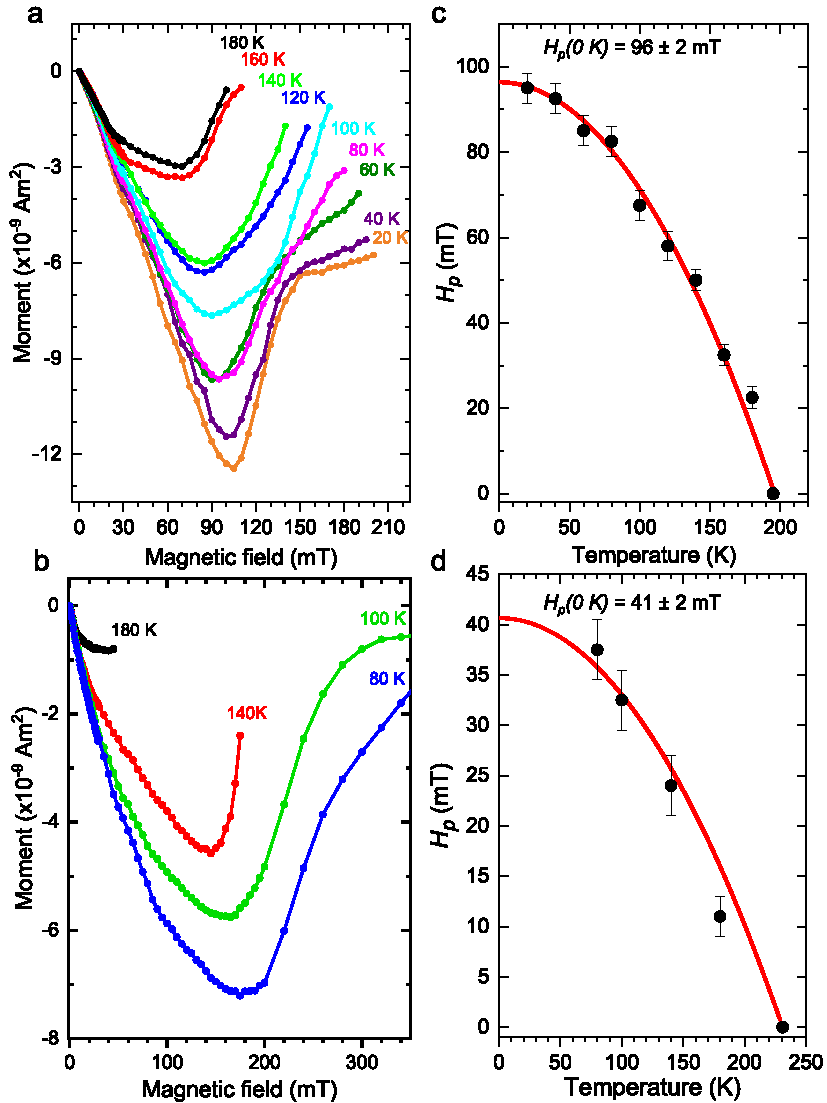}}
		\caption{Figs. 3a-d from Ref. \cite{e2021p}. For an explanation see text.}
		\label{figure1}
	\end{figure}
	
	\section{Introduction}
	
	In Figs 3(a) and (b) of Ref. \cite{e2021p}, reproduced here in the left panels
	of Fig. 1, the authors report the measured magnetic moment versus applied magnetic
	field for H$_{3}$S and LaH$_{10}$ samples at various temperatures. The curves show
	a diamagnetic response that is approximately linear up to a temperature-dependent
	field. That field was inferred to be the temperature-dependent lower critical
	field after applying a demagnetization correction. In Figs. 3(c) and 3(d) of
	\cite{e2021p}, reproduced here in Fig. 1 right panels, the authors plot this critical
	field versus temperature, showing behavior consistent with that expected for
	superconductors.
	
	\begin{figure}[t]
		\resizebox{8.5cm}{!}{\includegraphics[width=6cm]{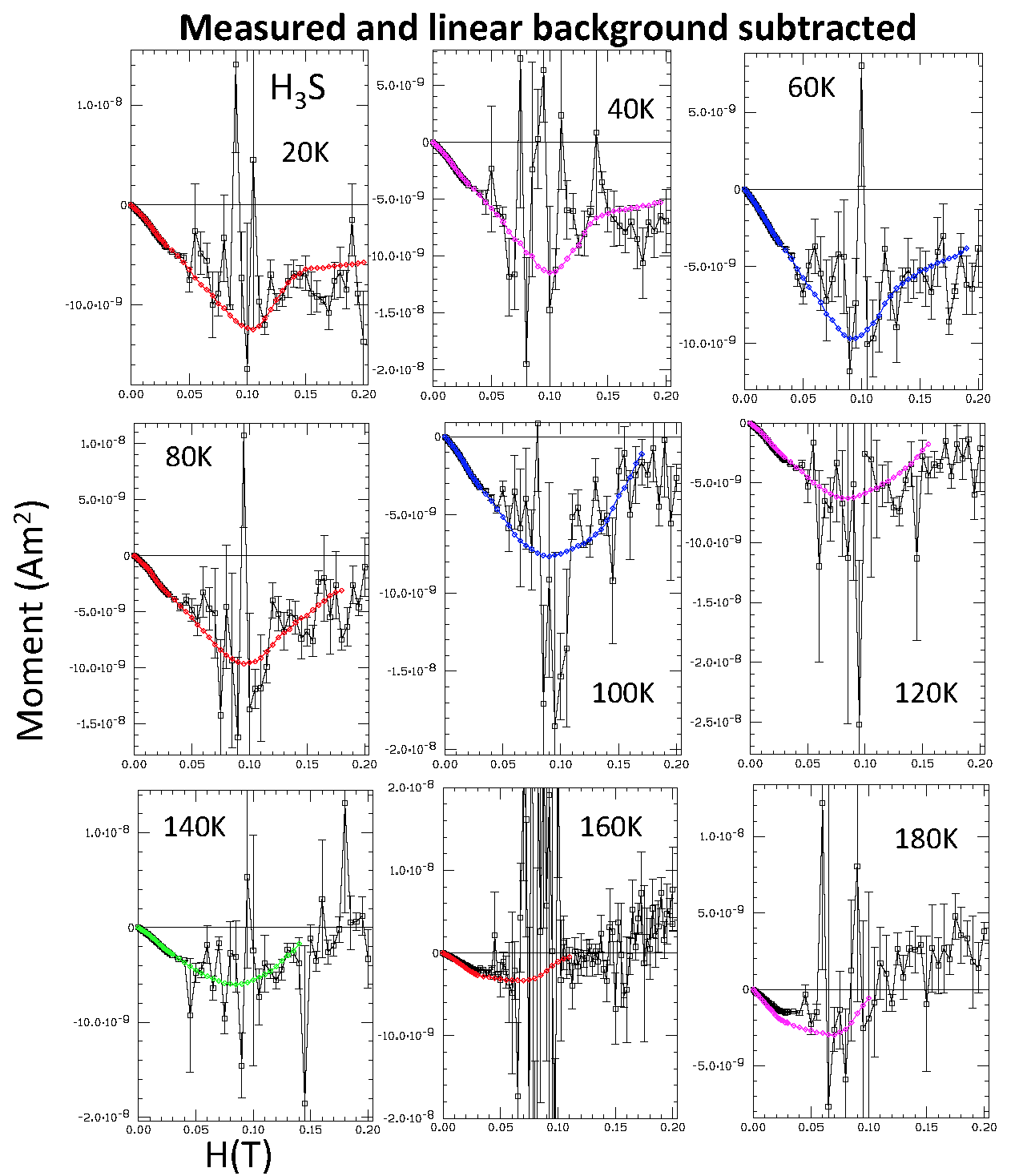}}
		\caption{ The 9 panels show the magnetic moment versus applied magnetic field
			obtained from the measured data reported in Ref. \cite{reval} for $H_{3}S$
			after subtraction of a linear background obtained from connecting the
			measured values on the virgin curves for field values 0T and 1T, as indicated
			in the Author Correction \cite{correction}. The colored curves on the 9
			panels show the corresponding curves obtained by digitization of the data
			points presented in Fig. 3a of \cite{e2021p} (reproduced here in Fig. 1a).}
		\label{figure2}
	\end{figure}
	
	In an Author Correction \cite{correction} prompted by questions raised in Ref.
	\cite{hm23}, the authors informed that their data in Figs. 3(a) and (b) were not
	directly measured, but instead were derived from the measured data by a
	variety of $linear$ transformations. Because the transformations were linear,
	the authors argued that they did not affect the values of critical field
	extracted from the point where the magnetic moment versus field started to
	deviate from linearity.
	
	In Refs. \cite{unpulling,bending, hysteresis,ma} we pointed out that it was mathematically
	impossible that some of the curves in Figs. 3(a) and (b) of \cite{e2021p} were
	derived by performing purely linear transformations on the reportedly measured
	data shown in Figs. 3(e), (f) and S10-12 of Ref. \cite{e2021p}. In their
	recently published revaluation paper \cite{reval} the authors inform that
	their measuring equipment
	{\it ``necessitated smoothing of a small portion of the data''}, without providing
	any procedure that can be used to reproduce the results. In a Nature News
	article \cite{garisto} that appeared at the same time the authors describe the
	published curves as {\it ``‘guide lines’ plotted over the averaged data''}.
	
	In their revaluation paper \cite{reval} the authors perform a new analysis, this
	time determining the lower critical field from the actually measured data. The
	authors state that the result of the re-analysis are consistent with the
	earlier analysis. They also make available for the first time numerical values
	for the measured data \cite{data}, which had been requested by one of us 19
	months ago.
	
	In this paper we argue that (i) the revaluation analysis \cite{reval} does not
	answer the questions raised in Refs. \cite{unpulling,bending,hysteresis,ma} regarding
	the relation between the measured data and the data published in Ref.
	\cite{e2021p}, (ii) the new analysis is flawed rendering its conclusions invalid,
	and (iii) there are serious questions about the consistency of a variety of
	the authors' statements with facts.
	
	\begin{figure}[t]
		\resizebox{8.5cm}{!}{\includegraphics[width=6cm]{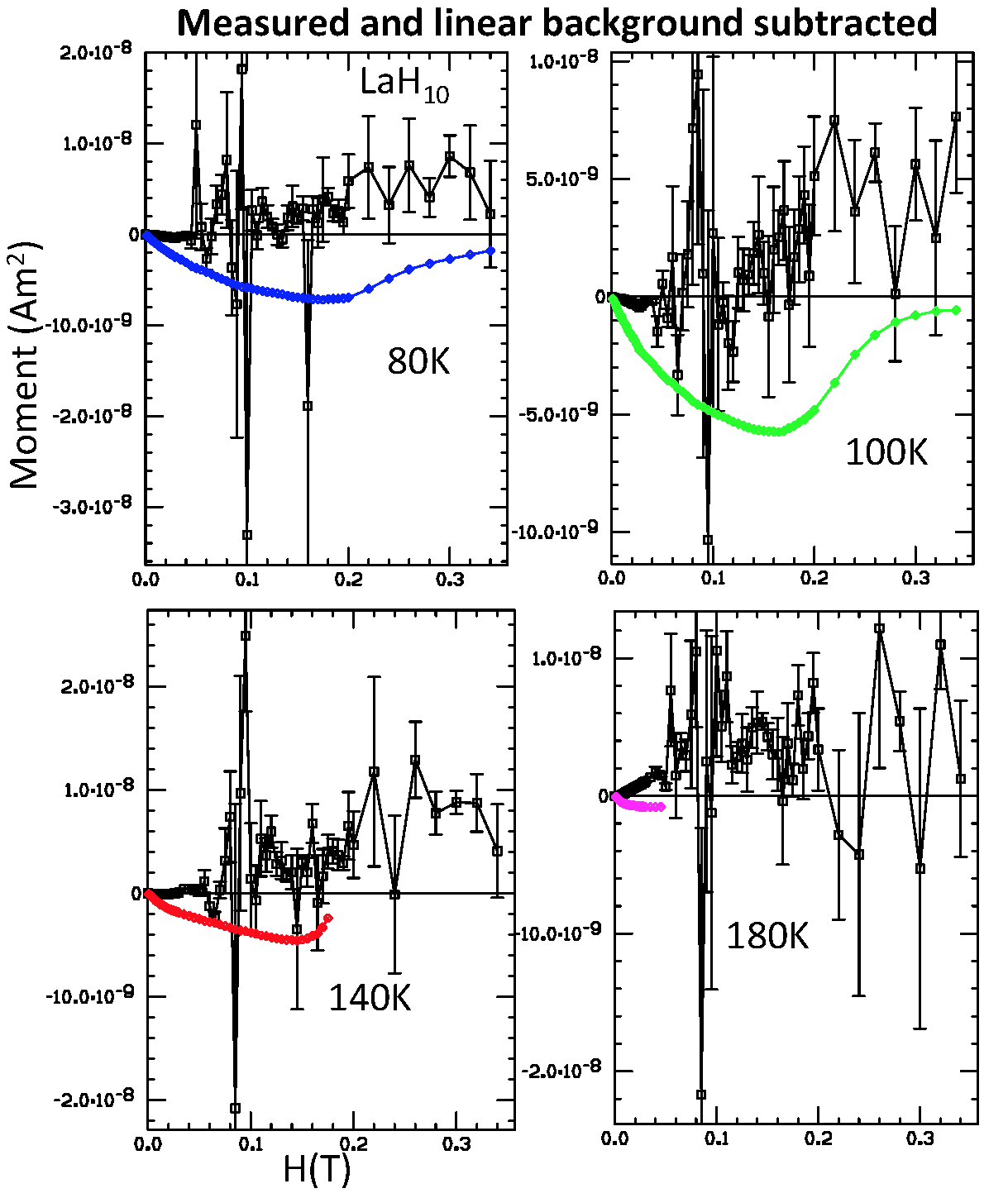}}
		\caption{The 4 panels show the results obtained from the measured data
			reported in Ref. \cite{reval} for $LaH_{10}$ after subtraction of a linear background
			obtained from connecting the measured values on the virgin curves for field values
			0T and 1T, as indicated in the Author Correction \cite{correction}. The colored
			curves on the 4 panels show the corresponding curves obtained by digitization
			of the data points presented in Fig. 3b of \cite{e2021p} (reproduced here in
			Fig. 1b). Note that the initial slopes of the points obtained from the
			measured data are different from those in the colored curves.}
		\label{figure3}
	\end{figure}
	
	The outline of our paper is as follows. In Sect. II, we analyze the relation between
	the newly supplied data \cite{data} and the curves in Figs. 3(a) and (b) of Ref.
	\cite{e2021p}, and conclude that the curves and data points in Figs. 3a, b cannot
	be derived from the measured data \cite{data} in a way that is consistent with
	accepted scientific practice. We also note that the newly released data
	\cite{data} are incompatible with the data reported as measured data in Figs. 3e,
	3f and S10-12 of Ref. \cite{e2021p}. In Sect. III we analyze the identification
	of signal and background contributions to the measured data \cite{data} and what
	it says regarding possible superconductivity in the samples. In Sect. IV we address
	the new fitting procedure presented in the revaluation paper \cite{reval} and argue
	that it has a variety of flaws, rendering the determination of critical fields
	performed by the authors invalid. In Sect. V we summarize our conclusions.
	
	\begin{figure}[t]
		\resizebox{8.5cm}{!}{\includegraphics[width=6cm]{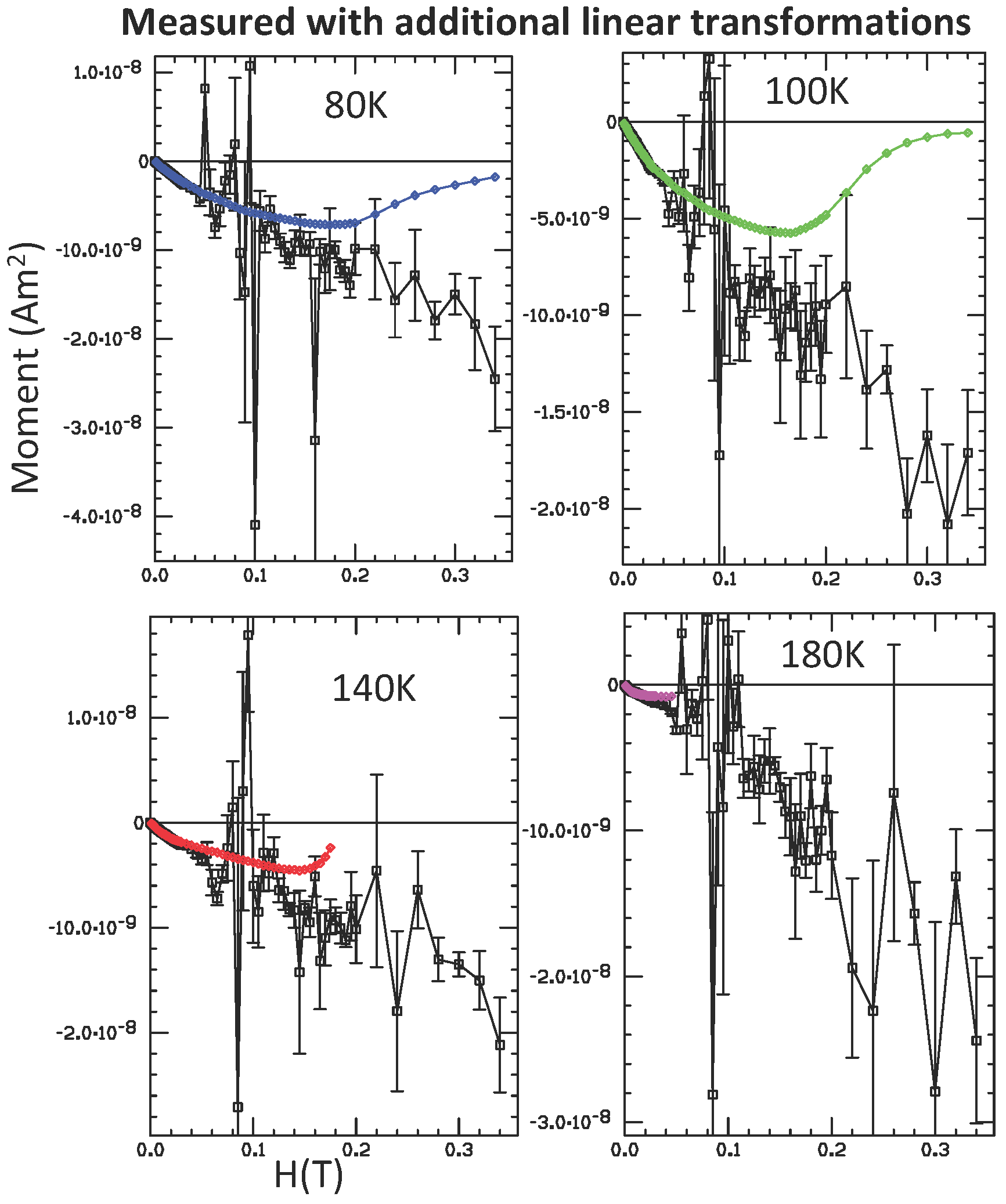}}
		\caption{Same as Fig. \ref{figure3} with additional linear background subtraction
			on the data in the 4 panels so that the initial slopes of the curves approximately
			match that of the colored curves obtained from Fig. 3b of \cite{e2021p}. Slopes
			of the lines subtracted from the measured data are $-4.56\times 10^{-7}Am^{2}
			/T$, $-4.64\times 10^{-7}Am^{2}/T$, $-4.20\times 10^{-7}Am^{2}/T$ and $-4.50\times
			10^{-7}Am^{2}/T$ for T=80K, 100K, 140K, 180K respectively.}
		\label{figure4}
	\end{figure}
	
	\begin{figure}[t]
		\resizebox{8.5cm}{!}{\includegraphics[width=6cm]{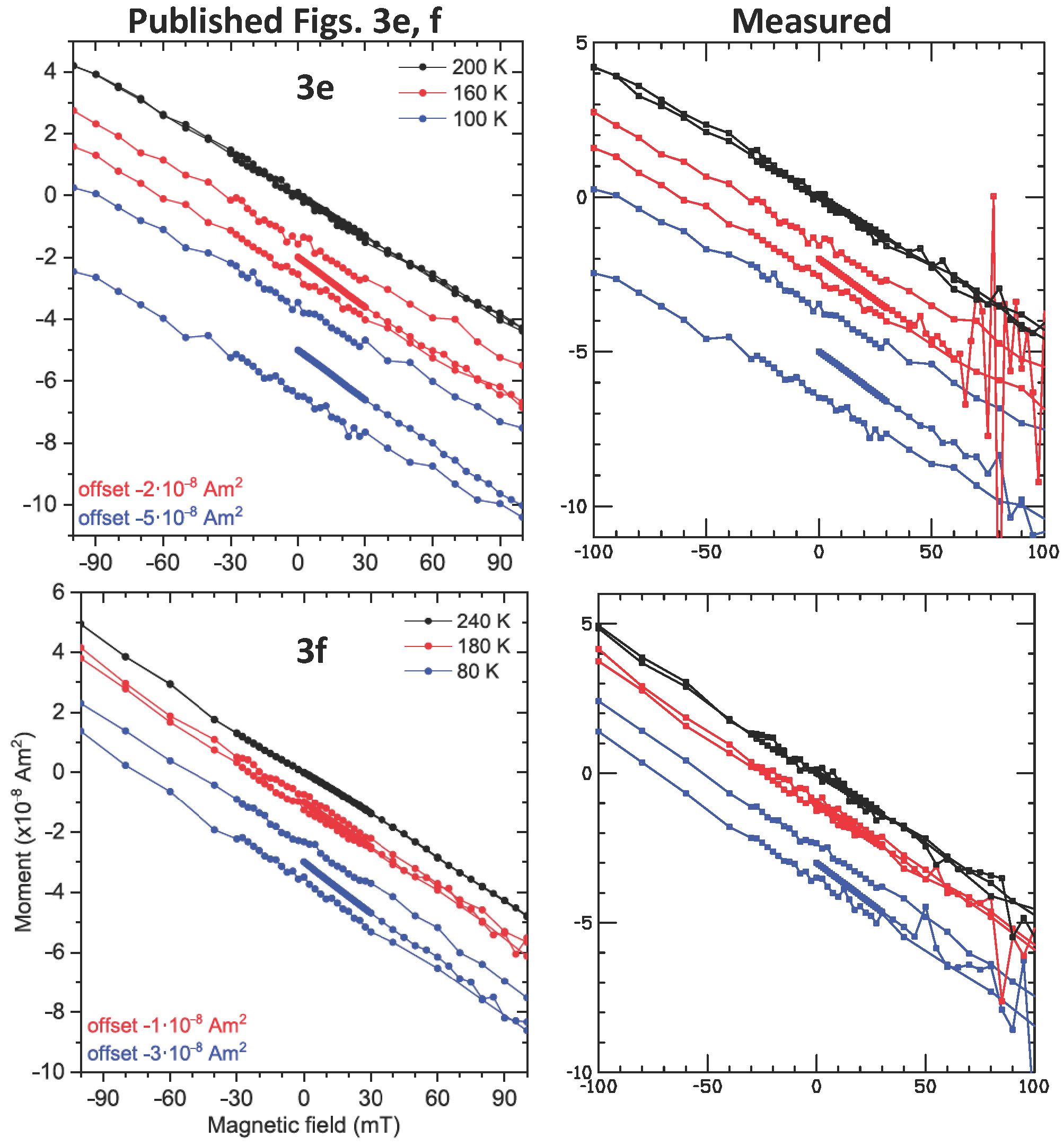}}
		\caption{Left panels: Figs. 3e and 3f from Ref. \cite{e2021p}, nominally ``raw
			magnetization'' data for $H_{3}S$ (upper left panel) and $LaH_{10}$ (lower left
			panel) in the field range $-100mT<H<100mT$. The curves were vertically displaced
			by the offsets indicated in the panels for display purposes. The central
			curve for each color is the virgin curve. The right panels show the corresponding
			measured data recently reported \cite{data} with the same offsets. Note the
			differences between the right and left panels. }
		\label{figure5}
	\end{figure}
	
	\section{Comparison between measured data and published data}
	\label{sec:data}
	
	Fig. \ref{figure2} shows the background-corrected magnetic moment of $H_{3}S$
	versus field as derived from the measured data published in OSF \cite{data} (black
	curves). The background is determined following Ref. \cite{correction}: a straight
	line connecting the measured magnetization at 0 and 1 T. It can be seen that
	below 30 mT the data extracted from measurements mostly coincide with the data
	reported in Fig. 3(a) of \cite{e2021p} (colored curves). However, at higher
	fields they strongly deviate and the measured data exhibit very large noise. This
	is explained in Ref. \cite{reval} by the fact that the current source used was
	different for the larger fields, resulting in more noisy data that had to be ``smoothened''
	to obtain the curves in Figs. 3(a) and (b) of Ref. \cite{e2021p}.
	
	Similarly, Fig. \ref{figure3} shows the measured magnetic moment versus field
	for $LaH_{10}$ (in black), again after applying the 0-1 T linear background subtraction
	procedure. The measured data again shows very large noise for fields larger
	than 30 mT. However, unlike the H$_{3}$S results in Fig. 2, here the
	background subtraction procedure results in $M(H)$ curves that strongly
	deviate from what was reported in Fig. 3(b) of \cite{e2021p} (colored curves).
	We conclude that the background subtracted by the authors to arrive at their Fig.
	3(b) was not the linear background connecting the points for fields 0T and 1T as
	was stated in Ref. \cite{correction}.
	
	Ref. \cite{correction} also states that in addition to this linear background subtraction
	{\it ``we performed additional linear transformations so that the curves have the same initial linear M(H) slope''}.
	It can be seen that none of the initial slopes of the data in Fig. 3 match the
	corresponding initial slopes of the colored curves. Nevertheless, to attempt to
	understand the way Fig. 3b of \cite{e2021p} was obtained starting from the
	measured data, we performed an additional linear transformation on the data for
	each the panels of Fig. 3 so that the initial slopes match the corresponding
	colored curves. The results are presented in Fig. 4: even ignoring the noise there
	seems little relation between the released data and the published curves.
	
	In addition to the question of how these ``additional linear transformations''
	were chosen to obtain the curves in Fig. 3(b) of \cite{e2021p}, another
	unanswered question is how the very scattered measured data for fields above
	30 mT in Figs. 2 and 4 collapsed to lie on the smooth curves shown in Figs. 2
	and 4. In \cite{reval} the authors refer to their Figs. 3a,b curves as being {\it ``smoothed data''}.
	We are unable to identify an algorithm or protocol consistent with accepted
	scientific practices that could yield the data points on the smooth curves in
	Figs. 2 and 4 starting from the measured data shown in those figures. It is
	for example unexpected that by using
	{\it ``multiple steps of smoothing/averaging''}, as reported in their revaluation
	paper \cite{reval}, the points in the field interval $60$mT$<$H$<100$mT in Fig.
	2 for T=20K, that all lie above the smoothed red curve, would collapse onto
	the points of that red curve. It is even more unexpected that ``smoothing'' the
	data in the top panels of Fig. 4 would yield the corresponding colored curves.
	
	Next we consider the relation of the data plotted in Figs. 3(e) and 3(f) of Ref.
	\cite{e2021p} with the measured data. In the Author Correction \cite{correction},
	the curves in Figs. 3(e) and 3(f) are referred to as {\it ``the raw magnetization curves''}.
	In Fig. 5 we reproduced the Fig. 3(e/f) curves from Ref. \cite{e2021p} in the
	left column, and the published measurement data in the right column. The scale
	of the panels is the same. While it is clear that many of the data points shown
	in the left and right columns are very similar, it is also clear that there
	are significant differences. In particular, for fields between 50 mT and 100
	mT the actually measured virgin curves show considerably more scatter than the
	'raw' data of Ref. \cite{e2021p} at the left. This suggests that the Fig. 3(e/f)
	data is smoothed, and thus is certainly not raw magnetization data. In
	addition, it can be seen that the as-published high temperature curves (200
	and 240 K) in black show no discernible hysteresis, as is expected for the non-superconducting
	state. In contrast, the measured data at the right show open loops for
	negative fields.
	
	\begin{figure}[t]
		\resizebox{8.5cm}{!}{\includegraphics[width=6cm]{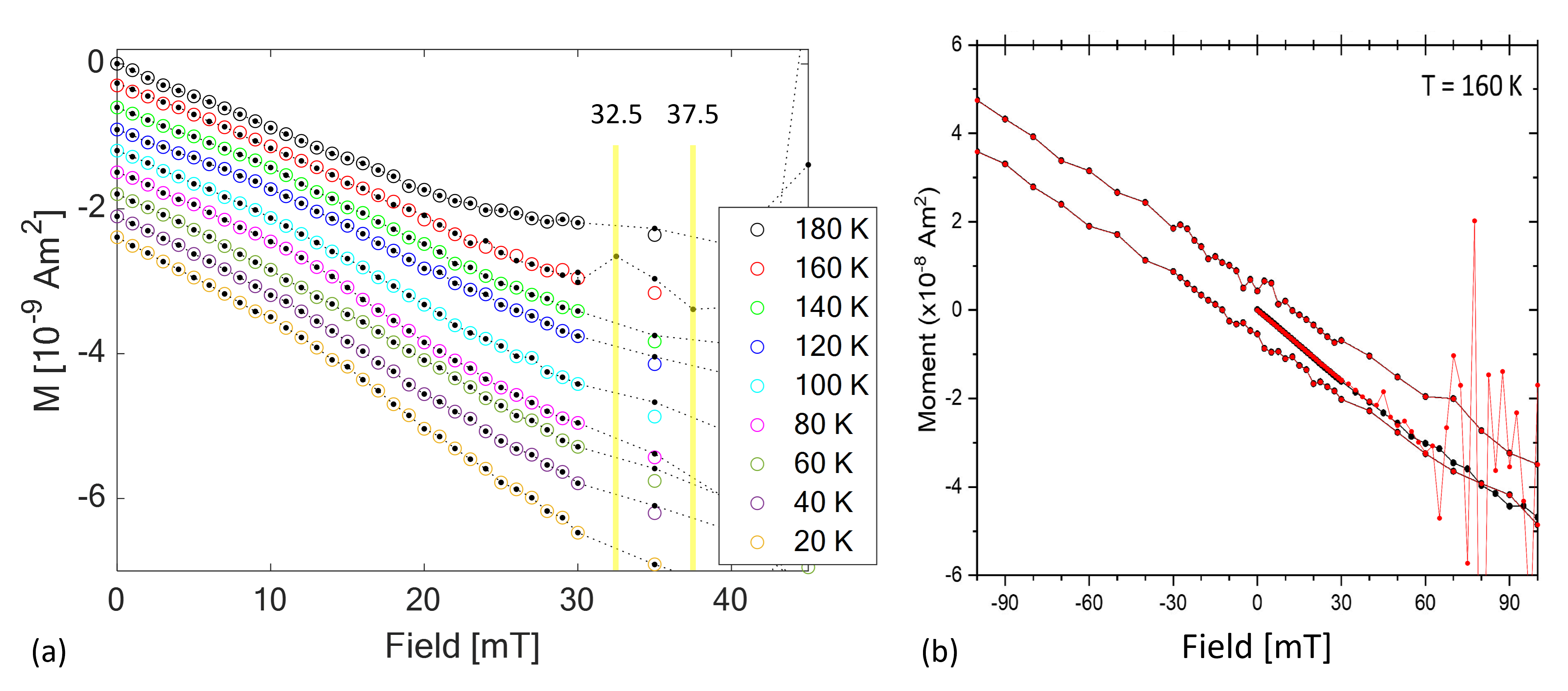}}
		\caption{ (a) Low field magnetization data for H$_{3}$S. Open symbols: as published
			in Ref. \cite{e2021p}, Fig. 3(a). Closed symbols: OSF dataset. Only the 160K
			datasets differs in the 0-30 mT range. Curves are vertically offset for clarity.
			(b) Comparison of the 160 K data in Fig. S10 of Ref. \cite{e2021p} to the
			OSF data. The virgin curve differs, but the hysteretic parts are identical. }
		\label{fig_datasets_h3s}
	\end{figure}
	
	\begin{figure}[t]
		\resizebox{8.5cm}{!}{\includegraphics[width=6cm]{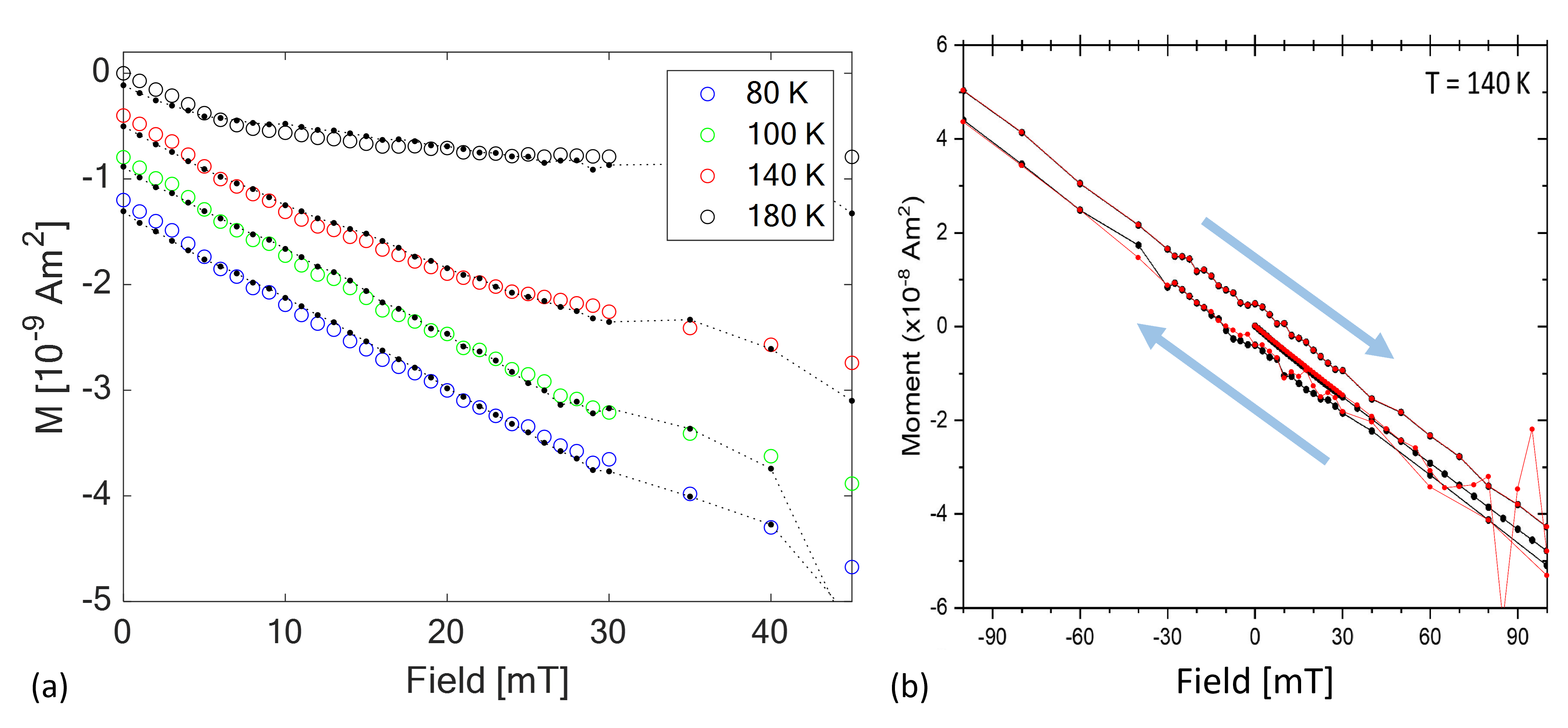}}
		\caption{ (a) Low field magnetization data for LaH$_{10}$S. Open symbols: as
			published in Ref. \cite{e2021p}, Fig. 3(b). Closed symbols: OSF dataset. The
			datasets do not match. Curves are vertically offset for clarity. (b)
			Comparison of the 140 K data in Fig. S11 of Ref. \cite{e2021p} to the OSF
			data. The virgin curve and the decreasing field hysteresis trace differ, but
			part of the increasing field hysteresis curve is identical. }
		\label{fig_datasets_lah10}
	\end{figure}
	
	The OSF dataset \cite{data} contains $M(H)$ traces for H$_{3}$S and LaH$_{10}$
	at temperatures that match those in the Nat. Comm. paper \cite{e2021p},
	suggesting a one to one relationship. However, it turns out that the actual
	correspondence is far from obvious. The comparison between the datasets is complicated
	by the apparent smoothing of the `raw' data published in the Nat. Comm. paper.
	However, it appears as if the authors did not smooth the $<30$ mT data, in
	line with the statement
	{\it ``we first smoothed the original data in the noisy range to combine it with the low-noise data''}
	in the revaluation paper. This part of the data can thus be used to identify datasets.
	
	In Fig. \ref{fig_datasets_h3s}(a) the low field H$_{3}$S magnetization is plotted
	for different temperatures, with the open symbols showing the Nat. Comm. data
	of Fig. 3(a) and the closed symbols the OSF data. The Nat. Comm. data was extracted
	from the published (vector) image with an accuracy of $\pm2\cdot10^{-11}$ Am$^{2}$.
	It can be seen that all but the 160 K datasets are identical at low fields.
	The 160 K datasets not only differ in their values, but also in field steps above
	$>30$ mT: 2.5 mT for the OSF data versus 5 mT the published curves. The yellow
	vertical lines show two of the additional field values. The actual relation is,
	however, more complicated. Panel (b) shows the magnetization data for the 160 K
	case on a wider field range. The curve in black is reproduced from Nat. Comm.
	Fig. S10, whilst the OSF data is shown in red. The center virgin curve is again
	seen to be different, especially for the 60-100 mT range where the Nat. Comm.
	curve shows no noise. The two hysteresis curves above and below are, however, identical.
	It thus appears that the two 160 K datasets are assembled from different
	pieces, sharing their hysteresis loop but using a different virgin curve.
	
	The same issues are found for the LaH$_{10}$ datasets, see Fig.
	\ref{fig_datasets_lah10}. The low field comparison in panel (a) shows that none
	of the virgin curves in the OSF dataset correspond to those published in the Nat.
	Comm. paper. However, looking at a wider field range one can again find matching
	stretches of data. Panel (b) compares the 140K $M(H)$ curves. The virgin curve
	and the decreasing field hysteresis curve are different. However, the increasing
	field curve \emph{is} identical in the range shown. The 140 K OSF dataset is stored
	in a single file with a timestamp for each datapoint. From that we learn that
	the OSF data was taken in one continuous 58 hour run. The 140 K Nat. Comm. dataset
	thus appears to be a mixture of different runs, seemingly swapping out \emph{half}
	a hysteresis loop and the virgin curve for other measurements.
	
\begin{figure}[t]
		\resizebox{8.5cm}{!}{\includegraphics[width=6cm]{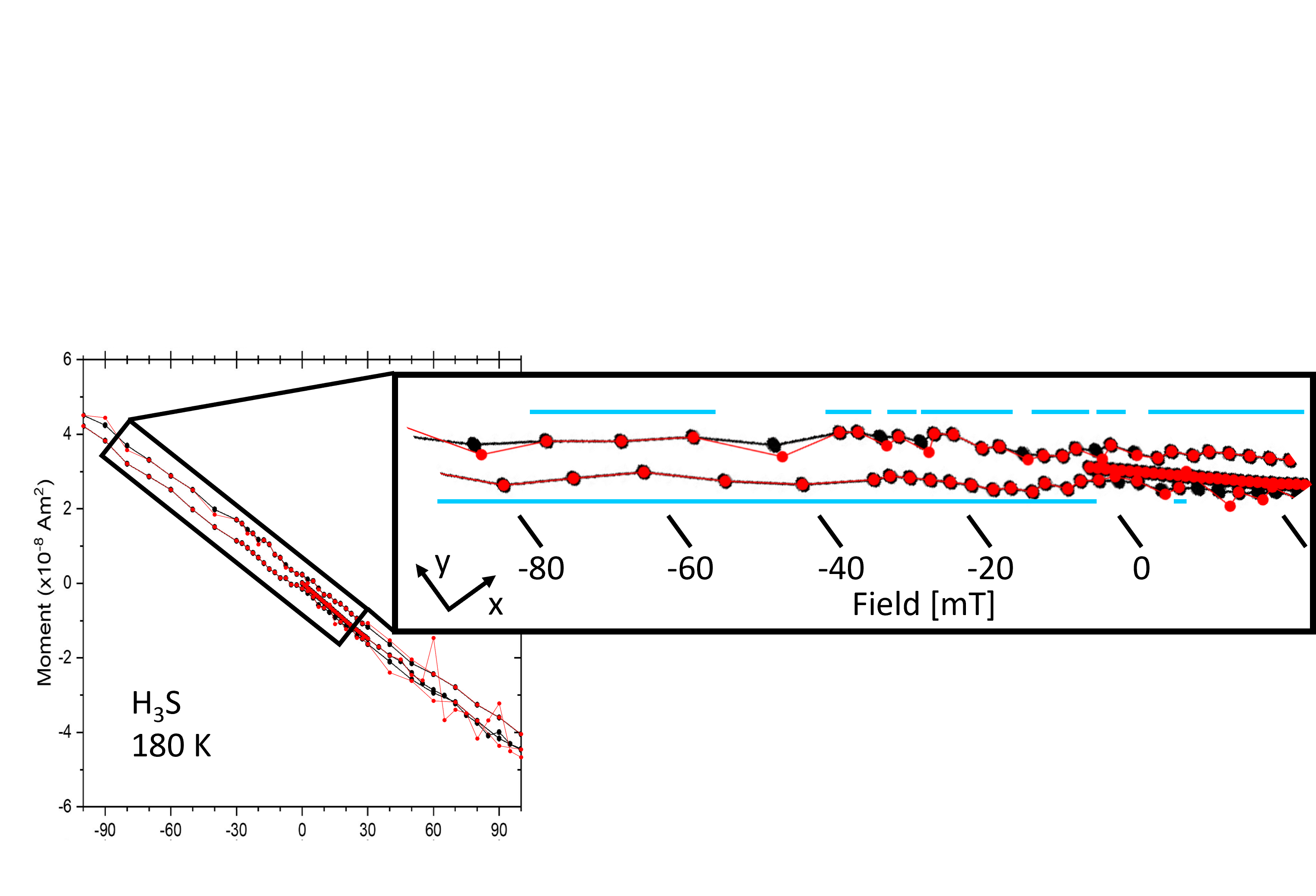}}
		\caption{ 180 K H$_{3}$S data from Fig. S10 in the Nat. Comm. paper (black)
			and the OSF data in red. The inset shows the -80...20 mT range rotated and
			enlarged. The blue lines show the intervals in which the datapoints match between
			the sets. }
		\label{fig_datasets_h3s_180K}
	\end{figure}

	The picture becomes even more confusing when looking at e.g. the H$_{3}$S 180
	K data, a continuous 77 hr measurement reproduced in Fig.
	\ref{fig_datasets_h3s_180K}. Its inset shows a zoom-in on the -80 to +20 mT range
	(rotated), with the blue lines showing datapoints that coincide between the
	Nat. Comm. (black) and OSF (red) data. For the lower hysteresis curve the two datasets
	match, but only for fields $\leq0$ mT; for positive fields the sets are different
	and increasingly so for higher fields. It is obviously no good practice to
	publish a hysteresis loop that is not measured in one go, but instead is assembled
	from disjoint parts. For the upper hysteresis curve we find stretches of
	matching data separated by non-matching points. In the -100 to 100 mT range we
	find that 28 of the 39 datapoints are indiscernible. The high proportion of matching
	datapoints excludes a coincidence: the upper hysteresis curve published in the
	Nat. Comm. paper must be related to data deposited on OSF. Each $M(H)$ point is
	the average of five z-scans. We checked whether the removal of outliers could
	be responsible for the observed differences, but found this not to be the case
	(see Fig. S3 in Supplementary Information section). The most likely explanation
	for the observed differences appears to be 'retouching' of individual datapoints
	to create the appearance of a smooth measurement curve. In Fig. S3(a) another
	clear retouching example is given for the 100 K H$_{3}$S $M(H)$ curve.
	
	In Figs. S1 and S2 we provide a comparison for all temperatures and materials,
	plotting the measured data on top of the 'raw' data published in Fig. S10 and S11
	of \cite{e2021p}. It can be clearly seen that all the virgin curves in the Nat.
	Comm. paper show much less noise in the 60-100 mT range, indicating undisclosed
	smoothing. Also at other fields the Nat. Comm. curves are often (very)
	different. This may in part be explained by the conjecture that the published curves
	are assembled from pieces of several repeated runs, whilst the deposited OSF
	$M(H)$ data just presents single runs. This conjecture can, however, not explain
	the 180K and 100K discrepancies discussed above. Also the observation that the
	published curves appear mostly smoother than the measured data over the full
	field range suggests that more smoothing/retouching has been applied.
	
	We conclude that at least some fraction of the published data reported as raw
	measurement data in Refs. \cite{e2021p,correction} are not measured data, but
	were obtained through undisclosed transformations.
	
	\section{Identification of the background signal and sample signal}
	
	\begin{figure}[]
		\resizebox{8.5cm}{!}{\includegraphics[width=6cm]{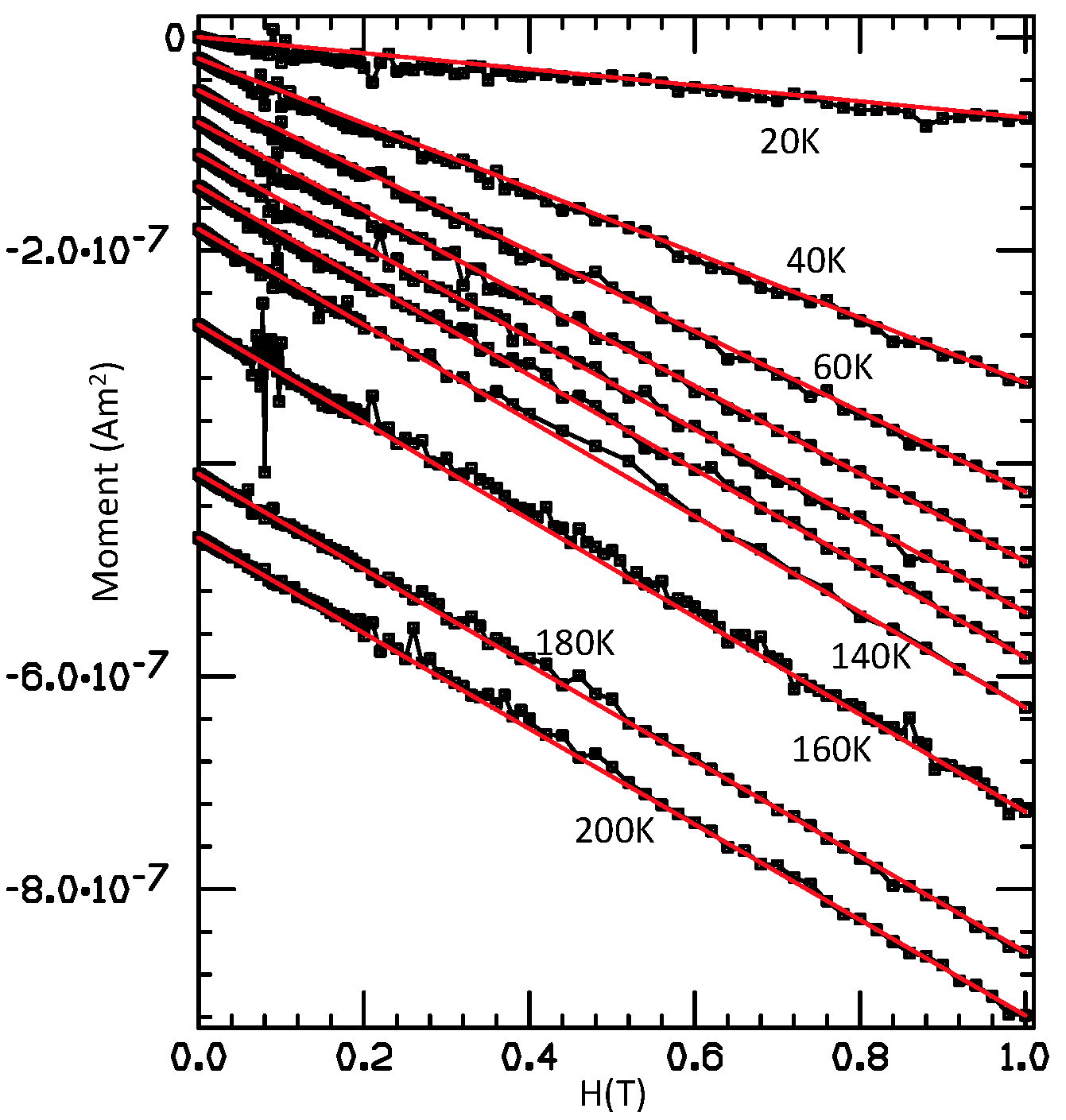}}
		\caption{ Magnetic moment versus magnetic field for $H_{3}S$. The straight
			red lines connect the magnetic moments at field zero and 1T and is assumed
			to be due to the background by the authors of Refs. \cite{e2021p,reval}. The
			difference between the points and the red line is assumed to be the magnetic
			moment of the sample which was plotted in Fig. 2 for small fields. The
			curves are shifted vertically for display purposes. Note the large difference
			in the slopes between the 20K and 40K data. }
		\label{figure6}
	\end{figure}
	
	All the data reported by the authors show a large diamagnetic response which is
	approximately linear in field. The authors assume that this response is the
	superposition of a strictly linear signal originating in the background and a
	much smaller diamagnetic response from the superconducting samples, that is linear
	for small fields and deviates from linearity above the lower critical magnetic
	field $H_{p}$ (corrected for demagnetization). Let us examine what the data
	say.
	
	\begin{figure}[]
		\resizebox{8.5cm}{!}{\includegraphics[width=6cm]{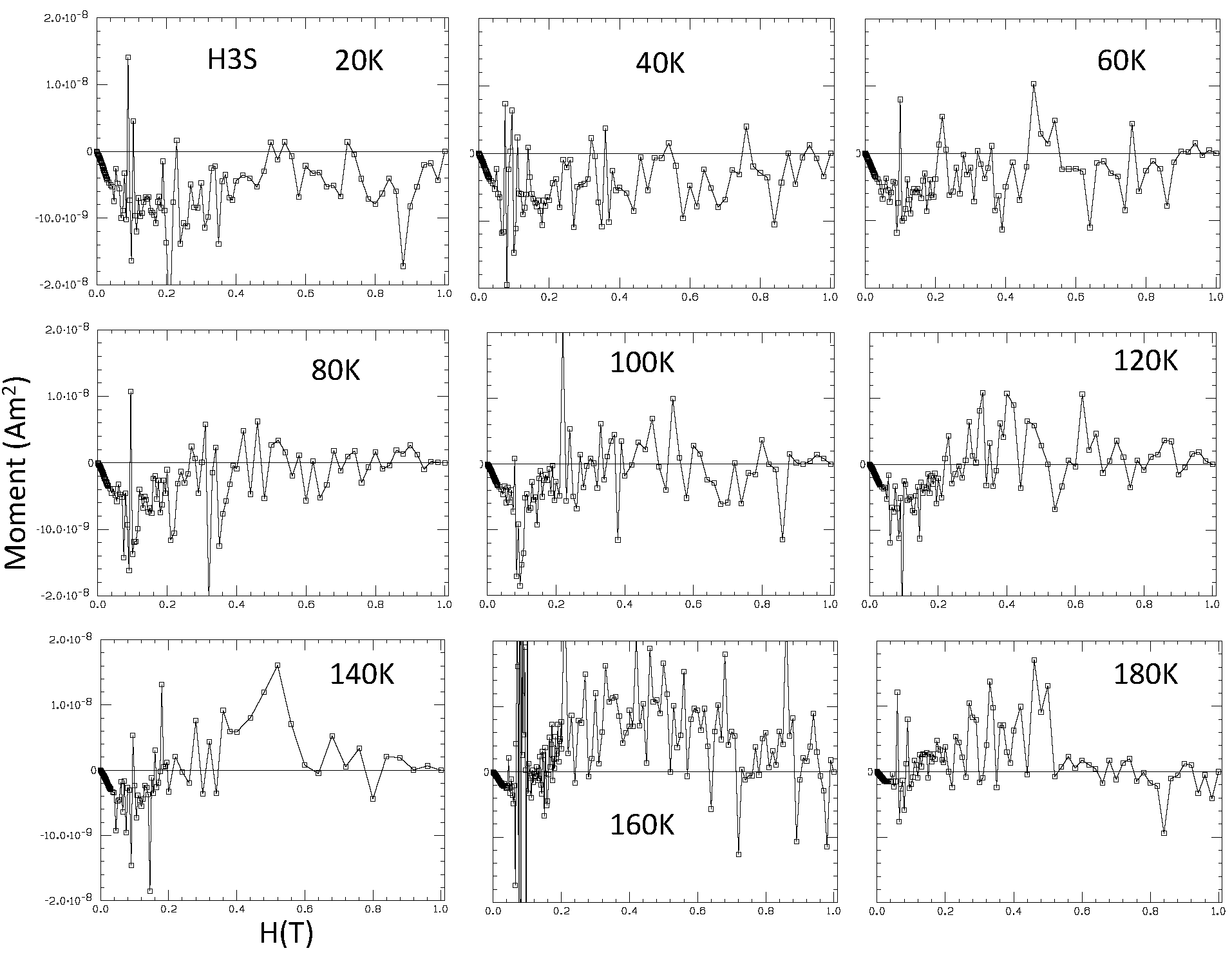}}
		\caption{ Magnetic moment versus magnetic field for $H_{3}S$ after subtraction
			of the linear function denoted by the red lines in Fig. 6, assumed to be the
			background signal by the authors of \cite{e2021p,reval}. }
		\label{figure7}
	\end{figure}

	In Fig. 9 we plot the raw data versus magnetic field for $H_{3}S$ and the
	straight red lines connecting the moments at 0T and 1T, assumed to be the background
	signal by the authors of Refs. \cite{e2021p,reval}. It can be seen,
	particularly for low temperatures, that the data points fall below the straight
	red lines for small fields, consistent with Fig. 2 and with the authors'
	interpretation that this may indicate the presence of a diamagnetic response
	of the sample itself, indicating superconductivity. In Fig. 10 we plot the same
	data with the background signal subtracted. According to the authors' interpretation,
	the data in Fig. 10 represent the moment of the superconducting sample. The
	critical temperature of this sample is assumed to be 195K \cite{reval}.
	
	This interpretation is called into question by the fact that, except at the lowest
	temperatures, the diamagnetic signal attributed to the sample disappears as
	the magnetic field increases, and the data points show a paramagnetic response
	for fields well below 1T and temperatures well below the assumed critical temperature.
	Instead, for a superconducting sample with very high upper critical field ($\sim
	90T$) as well as strong pinning centers, as is believed to be the case for
	these samples to be consistent with other experiments, one expects the diamagnetic
	response to persist well above magnetic field 1T for all temperatures smaller than
	the critical temperature. That expectation is not met by the data shown in Fig.
	10. If the slope of the subtracted background is assumed to be somewhat smaller
	than what the authors assumed the moment would turn upward at lower field and again
	downward at higher field, also not expected for a superconductor.
	
	The corresponding data for $LaH_{10}$ are shown in Figs. 11 and 12. Here it is
	seen that the response of the sample, if one assumes it to be the difference
	between the data points and the straight red line connecting the data points
	at field 0T and 1T, is paramagnetic for most of the temperature and field values.
	Even for very low fields the response of the sample is paramagnetic for
	temperatures T=140K and T=180K, well below the assumed critical temperature of
	this sample (231K), as seen in Fig. 12, as was also seen in Fig. 3. To obtain an
	initial response from the samples indicative of superconductivity, as was
	shown in Fig. 4 and as depicted by the authors in Fig. 3b of Ref. \cite{e2021p},
	it is necessary to assume that the background signal that has to be subtracted
	from the data to obtain the signal of the sample is given by the green lines
	in Fig. 8. However this would imply that the diamagnetic moment of the sample would
	grow very large for large fields, as seen in Fig. 4, inconsistent with the
	behavior expected for a small superconducting sample.
	%
	
	\begin{figure}[]
		\resizebox{8.5cm}{!}{\includegraphics[width=6cm]{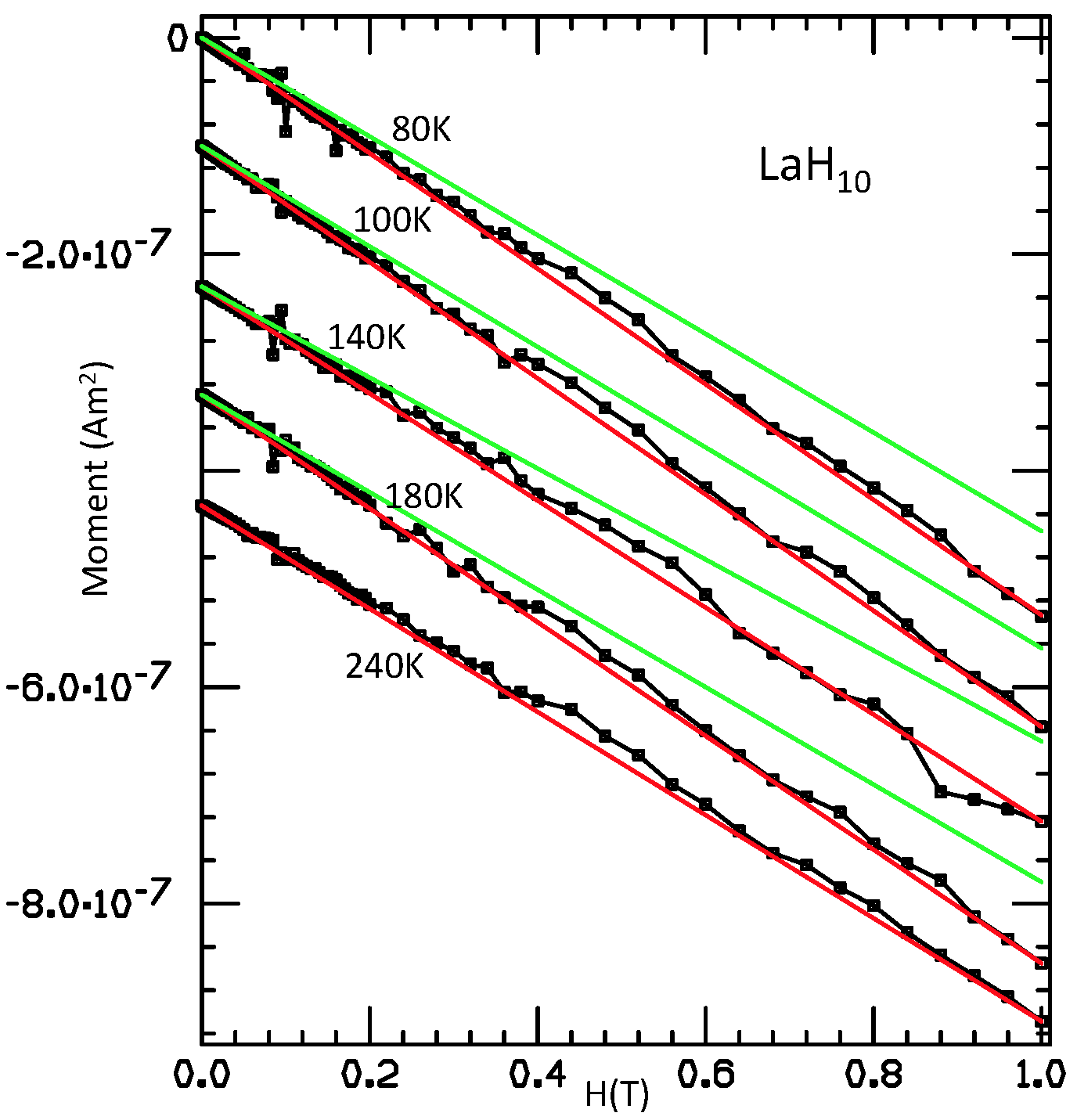}}
		\caption{ Magnetic moment versus magnetic field for $LaH_{10}$. The straight
			red lines connect the magnetic moments at field zero and 1T. The difference between
			the points and the red lines was plotted in Fig. 3 for small fields. The straight
			green lines are the linear background subtracted from the measured data to obtain
			the moments plotted in Fig. 4. The curves and lines are shifted vertically
			for display purposes. }
		\label{figure8}
	\end{figure}
	
	\begin{figure}[]
		\resizebox{8.5cm}{!}{\includegraphics[width=6cm]{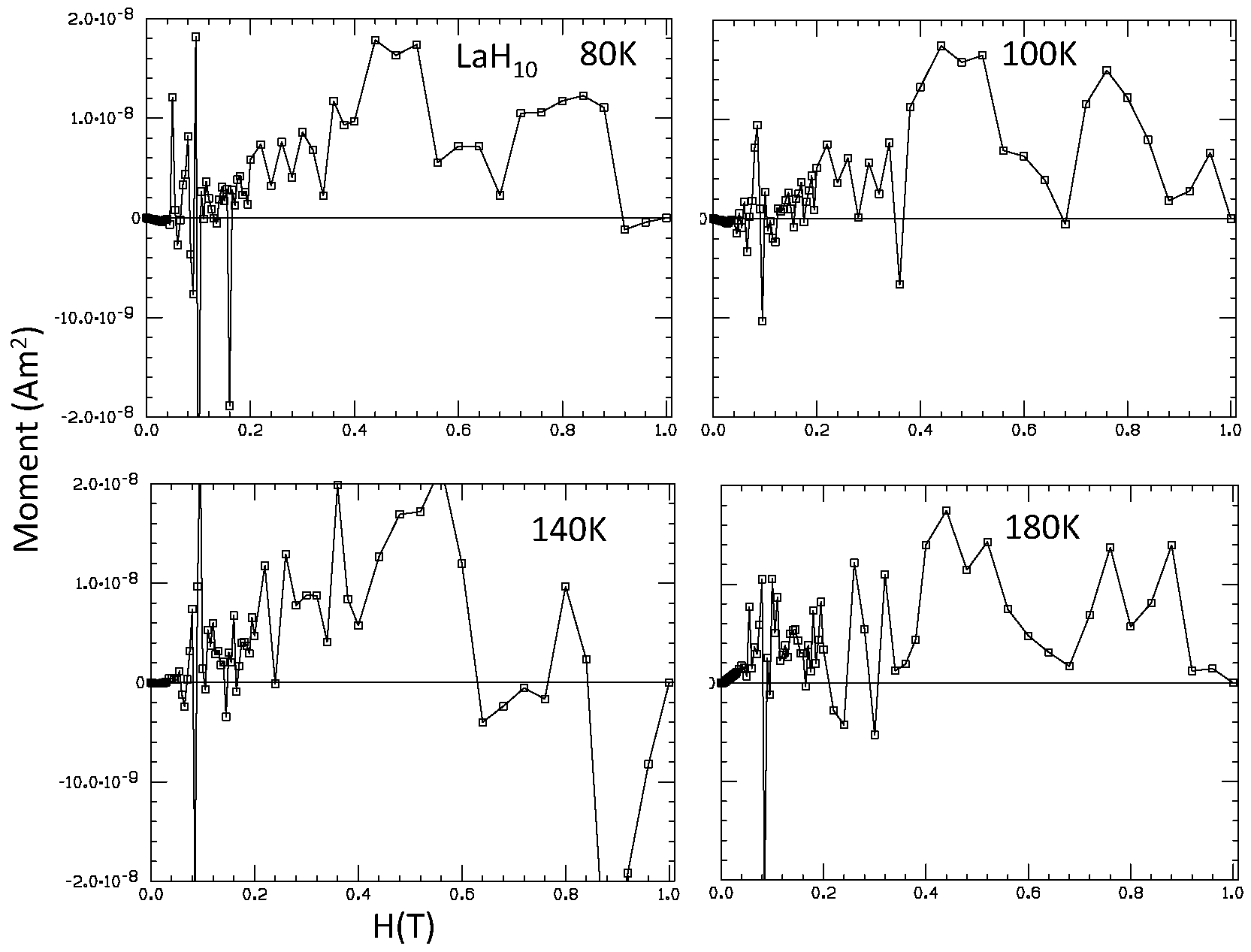}}
		\caption{ Magnetic moment versus magnetic field for $LaH_{10}$ after subtraction
			of the linear function denoted by the red lines in Fig. 8. }
		\label{figure9}
	\end{figure}
	
	These observations call into question the interpretation of the authors that
	what they identify as the part of the signal originating from the sample
	provides evidence that the samples are superconducting.

	\section{Determination of the lower critical field}
	
	In their original work \cite{e2021p,correction} the authors extracted the lower
	critical field $H_{p}(T)$ from the smooth $M(H, T)$ curves reproduced in Figs.
	1(a, b). $H_{p}$ is determined by eye from the point where the $M(H)$ curve deviates
	from linear. The relation between the published smooth $M(H)$ curves and the
	actual measured data is, however, lacking. This casts serious doubt on the validity
	of the $H_{p}(T)$ determination in \cite{e2021p}, see Fig. 1c,d above.
	
	To address this issue the authors recently published a revaluation of their lower
	critical field determination \cite{reval}. In the re-analysis the authors
	start from the raw measurement data \cite{data}. To tackle the issue of the large
	noise, a fitting procedure is used that is based on earlier work of Talantsev
	\cite{talan}: \beq \label{eq:fit} M(H) = m_0 + k\times H + m_c \times \left(
	\frac{H}{H_{p}} \right)^n \eeq The first two terms represent the linear
	contribution, the last term is an \textit{empirical} power law describing the
	deviation from linearity. The parameter $m_{c}$ is called the cut-off value:
	when the magnetic field $H$ reaches $H_{p}$, the magnetization deviates by $m_{c}$
	from linear. The parameters $m_{0}, k, H_{p}$ and $n$ are derived from the fit,
	whilst $m_{c}$ and the fitting range $0-H_{appl}$ are chosen by the authors.
	
	For a typical data point the authors perform five z-scans with their SQUID,
	reporting the average of these measurements as $M(H)$. The standard error of the
	mean (SEM) is used as a measure of the measurement uncertainty. When fitting
	the $M(H)$ data to Eq. \ref{eq:fit} the authors weight the errors, minimizing
	the function \beq \label{eq:chisq} \chi^2_{w} = \sum_{i} \left( \frac{M_{i}-M_{i,fit}}{SEM_{i}}
	\right)^2. \eeq
	
\begin{figure}[t]
		\resizebox{8.5cm}{!}{\includegraphics[width=6cm]{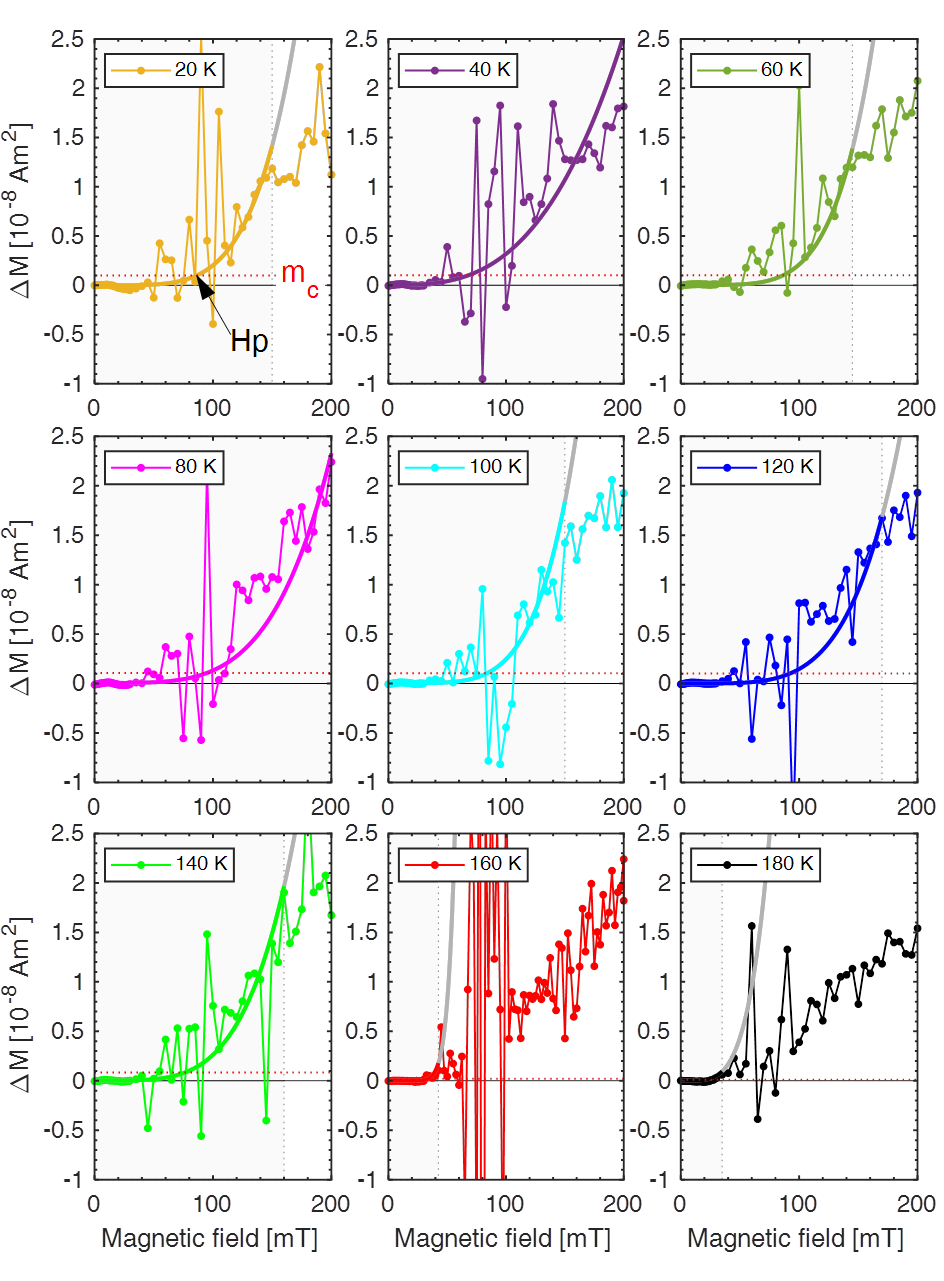}}
		\caption{ Magnetization versus field for H$_{3}$S at various temperatures together
			with fits to Eq. \ref{eq:fit} (smooth lines). The initial linear response has
			been subtracted from the raw $M(H)$ data for clarity. The field range for
			fitting is shown by a gray background and the variation in $m_{c}$ by a
			horizontal red line. $H_{p}$ corresponds to intersection of the fit with the
			red $m_{c}$ line. }
		\label{fig_deltam}
	\end{figure}

	Fig. \ref{fig_deltam} summarizes the fitting of the H$_{3}$S magnetization
	data at different temperatures. Each panel shows $\Delta M(H)$, the deviation
	of $M(H)$ from its initial linear $m_{0}+ k \times H$ behavior, together with the
	corresponding fit. The $0-H_{appl}$ fitting interval and $m_{c}$ vary from
	panel to panel, with the fitting interval indicated by the gray background and
	the cut-off value by a horizontal red line.
	
\begin{figure}[t]
		\resizebox{8.5cm}{!}{\includegraphics[width=6cm]{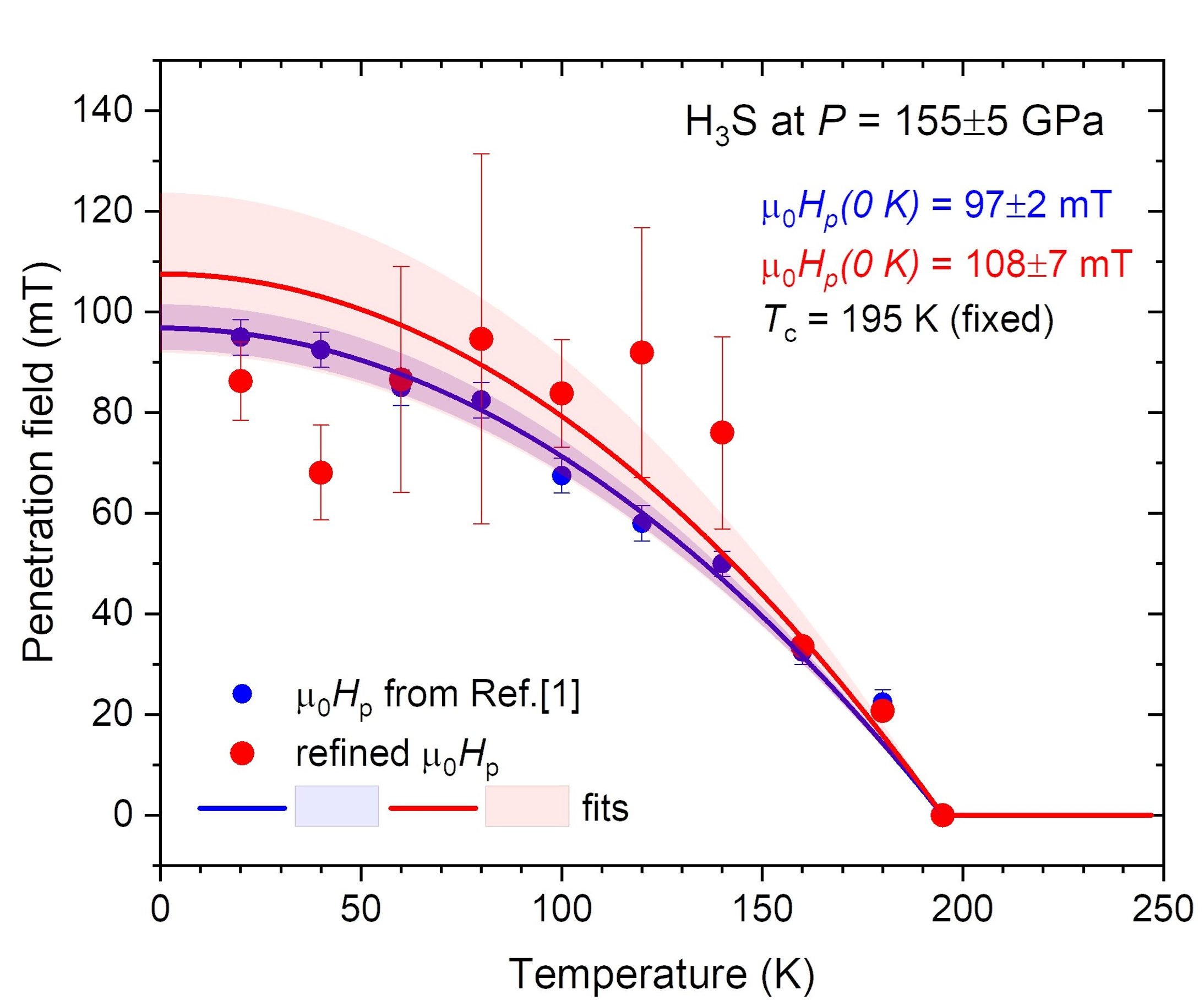}}
		\caption{ Fig. 5(a) from \cite{reval}. The blue data points and fit
			correspond to the by eye determination as published in \cite{e2021p}, the red
			points result from the fit-based re-evaluation. }
		\label{fig_hpt}
	\end{figure}

	The result of the authors' re-analysis is similar to that of the original
	analysis, see Fig. \ref{fig_hpt}. The re-analysis yields $H_{p}(0) = 108\pm7$ mT,
	which is within $\sim10 \%$ of the earlier reported value of $97\pm2$ mT. In the
	original analysis the $H_{p}(T)$ points (blue symbols) were tightly following a
	$1- \left(\sfrac{T}{T_c}\right)^{2}$ behavior (blue solid line). In the re-analysis
	this is no longer the case; the points have large error bars, are scattered, and
	in the 20-140 K range they could as well sit on a horizontal $\sim 80$ mT line.
	
	In the following we will take a critical look at the revaluation, in
	particular the choices made for $m_{c}$ and $H_{appl}$ and how they affect the
	results. In doing so Eq. \ref{eq:fit} is assumed to be a valid approximation
	of the low-field and virgin part of the $M(H)$ curves of hard superconductors
	as proposed in Ref. \cite{talan}.
	
	\subsection{Role of $m_{c}$}
	
	In his original publication \cite{talan} Talantsev proposed the following
	method to determine $m_{c}$: \beq m_c = f \times \max(|M(H, T_{lowest})|)
	\label{eq:mc_det} \eeq A single $m_{c}$ is used for all temperatures, fixed to
	a value that is a fraction $f$ of the maximum diamagnetic response of the superconducting
	sample. $f$ ranges from 0.005 (Eq. 11 in \cite{talan}) to 0.08 (Eq. 12), with the
	author stating that {\it``The best practice … is to utilize strict $m_{c}$ criterion defined by Eq. 11''}.
	Fig. 1a shows a maximum diamagnetic response of $1.2 \cdot 10^{-8}$ Am$^{2}$
	for H$_{3}$S. Using the $f=0.005$ criterion this would result in a fixed cut-off
	of $m_{c}= 6 \cdot 10^{-11}$ Am$^{2}$.
	
	In the revaluation paper \cite{reval} a very different approach is taken: $m_{c}$
	is not kept fixed, but is instead scaled with the uncertainty on the magnetic
	moment. For the 160 and 180K measurements the authors judge the low-noise $\le
	30$ mT points to be most important and state:
	{\it''we set the $m_{c}$ criterion as approximately three times the value of the uncertainty of the measured magnetic moment''}.
	This results in reported cut-off values of respectively $2\cdot10^{-10}$ and $1
	\cdot 10^{-10}$ Am$^{2}$. At lower temperatures also the noisy higher-field data
	is deemed important and another criterion is used:
	{\it ''we set the $m_{c}$ criterion to approximately the averaged value of uncertainties of the $m(H_{appl})$ datasets used for fitting''}.
	This choice leads to a $10\times$ higher cut-off value of $1\cdot 10^{-9}$ Am$^{2}$
	for the 20-120K data. However, in a nod to Eq. \ref{eq:mc_det} the authors assign
	a lower $m_{c}$ of $8 \cdot 10^{-10}$ Am$^{2}$ to the 140K dataset, as $m_{c}$
	would otherwise be {\it''larger than 10\% of the maximum diamagnetic response''}.
	We would like to note that in the original publication \cite{talan} the 10\% criterion
	is not mentioned. In fact, in its Sect. 3.7 a niobium $M(H, T)$ data set is
	fitted using a fixed $m_{c}$ that reaches up to 30\% of $\max(|M(H)|)$. Also, already
	the 100 K curve exceeds this criterion. According to Fig. 1a it shows a maximum
	diamagnetic response of $7.7\cdot10^{-9}$ Am$^{2}$, and hence its $m_{c}$ of $1
	\cdot10^{-9}$ Am$^{2}$ sits at 13\% of that response.
	
	Changing $m_{c}$ does not change the fitted curve in any way, but it has a
	direct effect on the value of $H_{p}$. It can be trivially shown that the following
	relation holds: \beq H_{p2} = (m_{c2}/m_{c1})^{1/n} H_{p1} \eeq with $H_{p1}$
	the field determined with a cut-off $m_{c1}$, and $H_{p2}$ the field that would
	follow when using a cut-off $m_{c2}$. Choosing a larger $m_{c}$ will thus
	directly result in a higher (reported) $H_{p}$ value. In Fig. \ref{fig_deltam}
	the effect of changing $m_{c}$ can be directly judged by vertically shifting
	the red $m_{c}$ lines.
	
	\begin{figure}[t]
		\resizebox{8.5cm}{!}{\includegraphics[width=6cm]{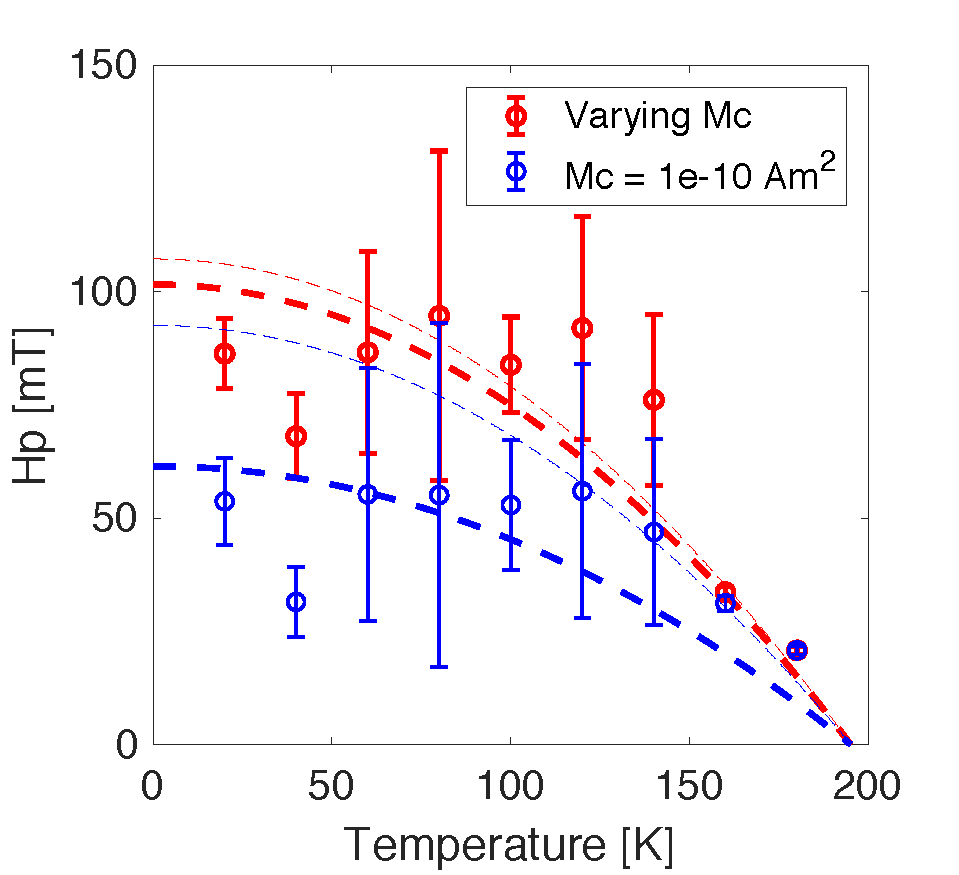}}
		\caption{ $H_{p}$ versus temperature as determined by the authors with varying
			$m_{c}$ in red, and with a fixed $m_{c}$ of $1\cdot 10^{-10}$ Am$^{2}$ in blue.
			The thick dashed lines are unweighted fits to
			$H_{p}(0) \left[ 1- \left(\sfrac{T}{T_c}\right)^{2}\right]$, the thin lines weighted
			fits. }
		\label{fig_hp_mc}
	\end{figure}
	
	The authors provide no rationale for their scaling of the cut-off value with the
	(average) uncertainty on magnetic moment measurements, except that they
	believe that {\it''the deviation from the linear trend becomes reliable''}
	above the chosen cut-off value. However, scaling $m_{c}$ with noise level
	leads to illogical results. Take for example the analysis of a number of $M(H)$
	curves that have the exact same shape, but a different noise level. Such a situation
	can for example arise by measuring the same sample with a SQUID using a different
	number of Z-scans and hence averages. By scaling $m_{c}$ with noise level one
	will systematically find larger $H_{p}$ value for the noisier curves, despite them
	being identical in shape. The original $m_{c}$ choice, Eq. \ref{eq:mc_det}, does
	not suffer from this bias.
	
	In their re-analysis the authors use a 10x larger $m_{c}$ value for the low temperature
	curves, resulting in significantly higher $H_{p}$ values at low temperatures. The
	effect of this choice is illustrated in Fig. \ref{fig_hp_mc}. The red points correspond
	to the authors’ choice of $m_{c}$, whilst for the the blue points a fixed
	$m_{c}$ of $1 \cdot 10^{-10}$ Am$^{2}$ is used. The latter number is equal to
	the authors’ 180 K $m_{c}$ value and corresponds to $f \approx 0.008$ (see Eq.
	\ref{eq:mc_det}). The constant $m_{c}$ choice results in a ~40\% reduction of
	the low temperature $H_{p}$ values \emph{without changing the fit curves}. At
	the same time $H_{p}(0)$ changes from $102 \pm 10$ to $61 \pm 8$ mT \footnote{
		\samepage Using an unweighted fit of $H_{p}(T)$. The authors use a weighted
		fit that assigns $10\times$ more weight to the 160 and 180K data points, which
		thereby dominate the fit. The weighted choice results in a ~15\% reduction of
		$H_{p}(0)$}. The latter value is obviously corresponding less well with the originally
	reported $H_{p}(0) = 97 \pm 2$ Am$^{2}$. It is important to note that the freedom
	in choosing $m_{c}$ is not reflected in the error on the reported $H_{p}(0)$ values.
	
	In their revaluation the authors deviated from Ref. \cite{talan} in their
	determination of $m_{c}$. Using their alternative method the newly determined $H
	_{p}(0)$ is consistent with that of the original publication. At the same time
	the alternative method appears to be flawed, resulting in larger reported $H_{p}$
	values for noisier datasets. In absence of a well-substantiated method for its
	determination $m_{c}$ is just a tuning parameter whose uncertainty is not
	reflected in the reported $H_{p}(T)$ values.
	
	\subsection{Role of the fitting interval}
	
	\begin{figure}[t]
		\resizebox{8.5cm}{!}{\includegraphics[width=6cm]{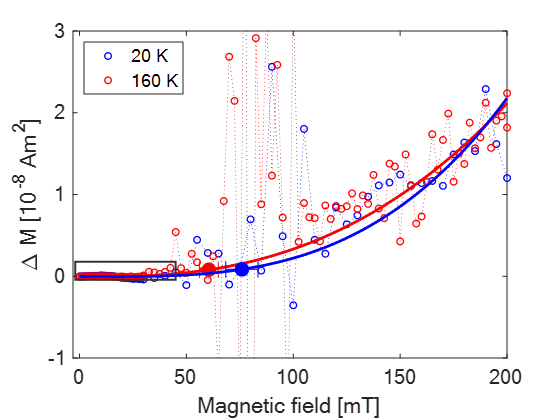}}
		\caption{ The 20K (blue) and 160K (red) magnetization curves are very similar.
			The solid lines are fits using the authors' 20 K $H_{appl}$ and $m_{c}$ choices
			(150 mT, $m_{c}= 1 \cdot 10^{-9}$ Am$^{2}$). The solid symbols show the
			resulting $H_{p}$ and its error. The gray rectangle indicates the area shown
			in Fig. S1 of the revaluation paper for the 160K fit. }
		\label{fig_fitrange}
	\end{figure}
	
	\begin{figure}[t]
		\resizebox{8.5cm}{!}{\includegraphics[width=6cm]{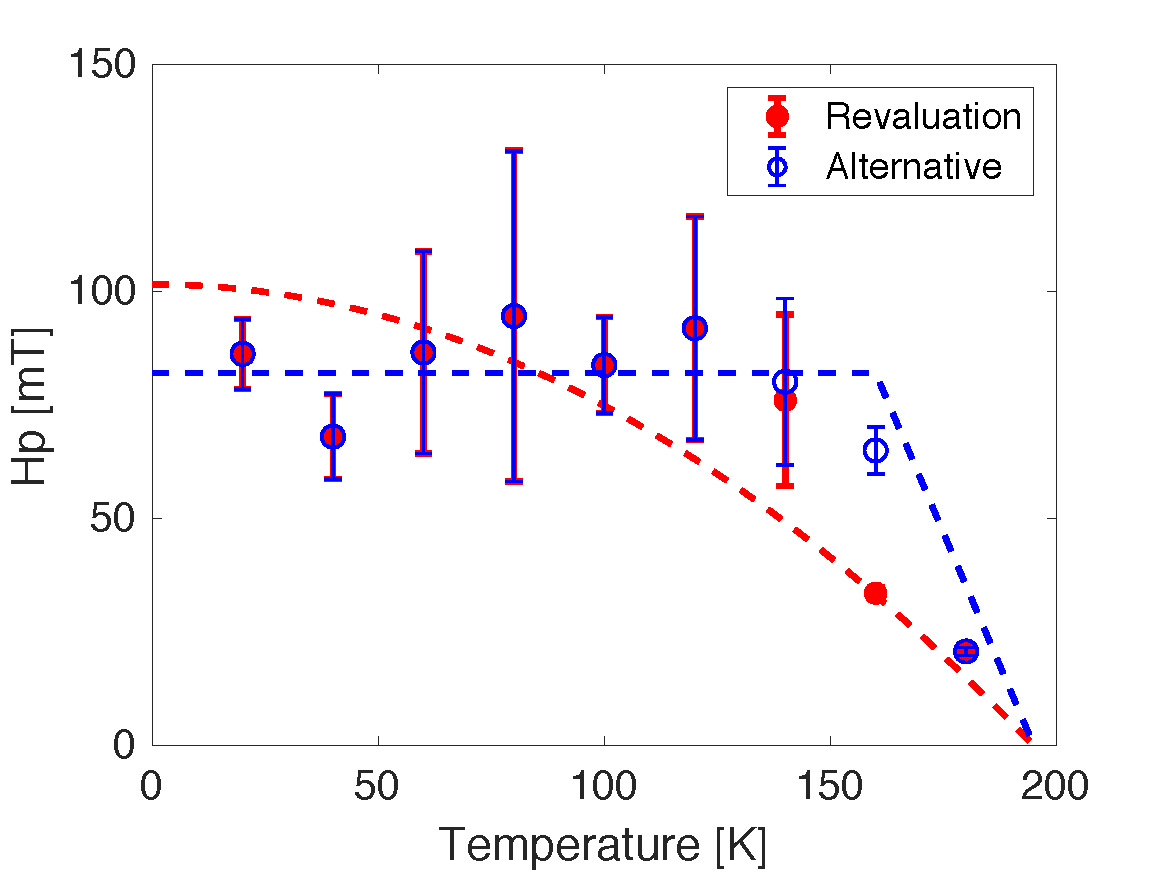}}
		\caption{ Effect of fit range on $H_{p}(T)$ interpretation. Red: authors'
			choice of fit range and $m_{c}$ with the dashed line a fit to
			$H_{p}(0) \left(1-\left( \sfrac{T}{T_{c}}\right)^{2}\right)$. Blue: Same, but
			using a 0-150 mT fitting range for the 160 K point and $m_{c}=10^{-9}$ Am$^{2}$
			for 140 and 160K. Dashed line is a guide line through the data. }
		\label{fig_alternative_hp}
	\end{figure}
	
	The empirical power law of Eq. \ref{eq:fit} is a simple approximation of the
	actual $M(H)$ curve and it is therefore only valid over a limited field range.
	In their revaluation the authors limit the range of $M(H)$ data used for fitting
	to the value of $H_{appl}$ at which the difference between the measured $M(H_{appl}
	)$ and the linear term reaches approximately $10 \times m_{c}$. It may be that
	this criterion is behind the authors’ choice to scale $m_{c}$ with the noise
	level. For noisy data one needs to ‘average’ over more points, and hence a
	larger field range, to get a reasonable result; increasing $m_{c}$ in
	combination with the above criterion does exactly this. Similarly, reducing
	$m_{c}$ at higher temperatures limits the field range used for fitting and
	hence allows one to exclude the noisier measurements. It needs no saying that in
	reality the approximation of Eq. \ref{eq:fit} \emph{does not} become valid over
	a larger field range when the noise or cut-off $m_{c}$ is larger.
	
	The choice of fitting interval can have a significant effect on the resulting fit
	curve. Take for example the 160 K H$_{3}$S fit as shown in the bottom-middle
	panel of Fig. \ref{fig_deltam}. Already on that scale it is clear that the fit
	goes steeply up on what could well be a few outlying $\Delta M(H)$ points. At the
	same time the overall shape of the 160 K curve is not different from any of those
	at lower temperatures. In Fig. \ref{fig_fitrange} the 160K $\Delta M(H)$ curve
	is plotted together with the 20K curve, showing them to be near-identical in
	shape and magnitude over the plotted range. The gray rectangle indicates the view
	as presented in Ref. \cite{reval}. Given the similarity it makes sense to fit the
	160K curve with the same parameters as used for the 20K curve: $H_{appl}=150$
	mT and $m_{c}= 1\cdot10^{-9}$ Am$^{2}$. This then results in a $H_{p}$ of
	$65\pm5$ mT, nearly double the reported value.
	
	The blue points in Fig. \ref{fig_alternative_hp} show the effect of the
	revaluation of the 160 K fit on $H_{p}(T)$. In the authors' determination of
	$H_{p}$ (red symbols) the two lower values at 160 and 180K gave something of a
	semblance of $1- \left(\sfrac{T}{T_c}\right)^{2}$ scaling. With the revised 160
	K point a constant, $\sim80$ mT $H_{p}$ up to 160 K followed by a steep drop
	seems more fitting.
	
	Making a coupling between the fitting interval and the noise in the $M(H)$
	data is questionable. Even with this choice the determination of fitting interval
	is subjective, as it requires one to 'see' where $H_{p}$ is. For the 160K dataset
	a wider interval seems far more logical, resulting in an essentially flat
	$H_{p}(T)$ dependence.
	
	\subsection{Sub-optimal fits}
	
	When fitting the $M(H)$ data the authors minimize the weighted chi-square, Eq.
	\ref{eq:chisq}. For the LaH$_{10}$ most of the reported solutions appear sub-optimal.
	In Fig. 18a the 140 K $M(H)$ data is shown. The blue curve represents the
	authors' fit with a $\chi^{2}_{w}$ of $\sim 400$. The red curve is a refit, resulting
	in a lower $\chi^{2}_{w}$ of $\sim 100$. Choosing a different fitting interval
	does not reconcile the difference, suggesting that other constraints have been
	used. Fig. 18b shows a comparison of $H_{p}(T)$ as derived by the authors in red
	and by re-fitting in blue. Only the 80 K value is a match, both in value and error
	bar. The re-fitted $H_{p}(T)$ curve shows little correspondence with the
	$1-\left(\sfrac{T}{T_{c}}\right)^{2}$ fit.
	
	\begin{figure}[t]
		\resizebox{8.5cm}{!}{\includegraphics[width=6cm]{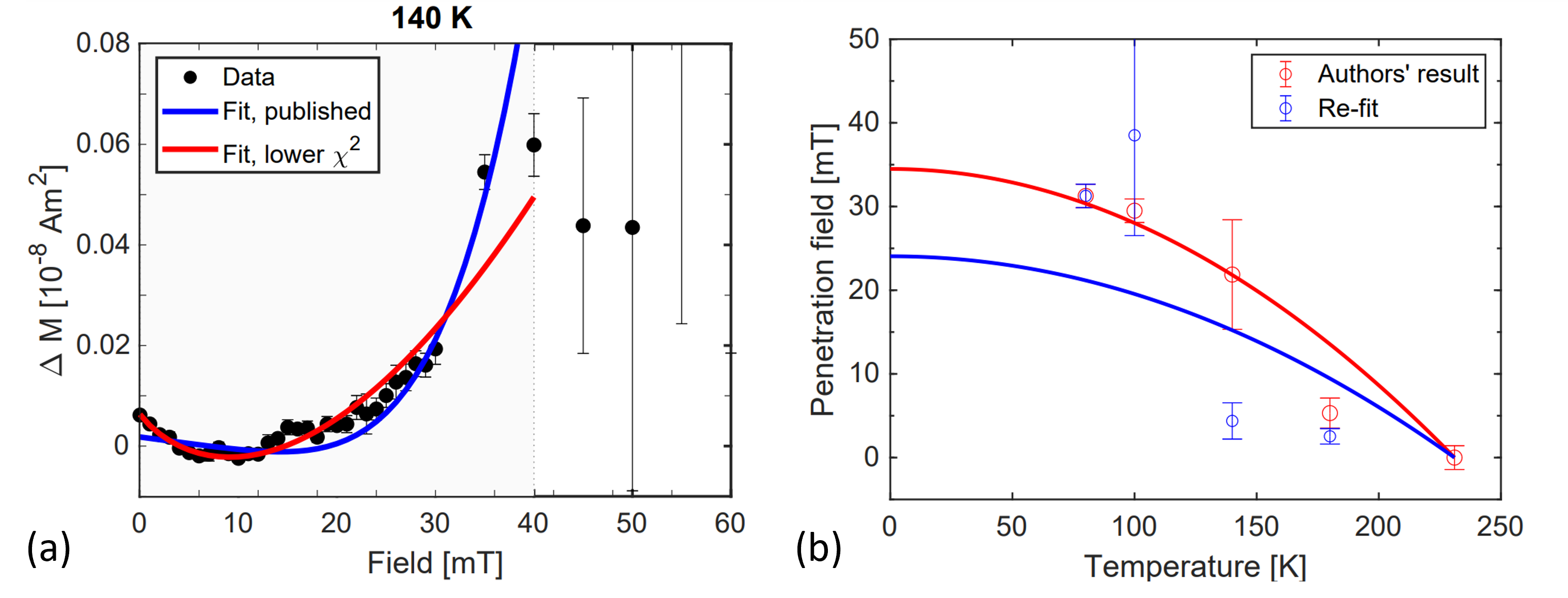}}
		\caption{ (a) LaH$_{10}$ 140 K data with the authors' fit in blue and a more
			optimal fit in red. The gray shaded area represents the 0-40 mT fit interval.
			(b) $H_{p}(T)$ as reported by the authors (red) and based on a re-fit of the
			data (blue). }
		\label{fig_misfits}
	\end{figure}
	
	\subsection{Mismatch in datasets}
	
	A revaluation can be expected to re-analyze the same data in a different
	manner. In Sect. II it was found that the data published in the OSF repository
	mostly does not match the the data published in \cite{e2021p}. For every dataset
	the virgin curve is different for fields $>30$ mT. This may be explained by
	undisclosed unconventional smoothing of the (raw) data shown in Figs. 3(e), 3(f),
	S10, S11, and S12 of the original publication and its correction. However, also
	many curves are found to differ in the low-noise and presumably unsmoothed $\leq
	30$ mT field range. For H$_{3}$S the 160K curve does not match, whilst for LaH$_{10}$
	there is no overlap at all between the virgin data analyzed in the original
	publication and in its revaluation.
	
	Presenting an analysis of mostly different datasets as a revaluation is a
	surprising choice.
	
	\subsection{Role of nonlinearity of the background}
	\begin{figure}[t]
		\resizebox{8.5cm}{!}{\includegraphics[width=6cm]{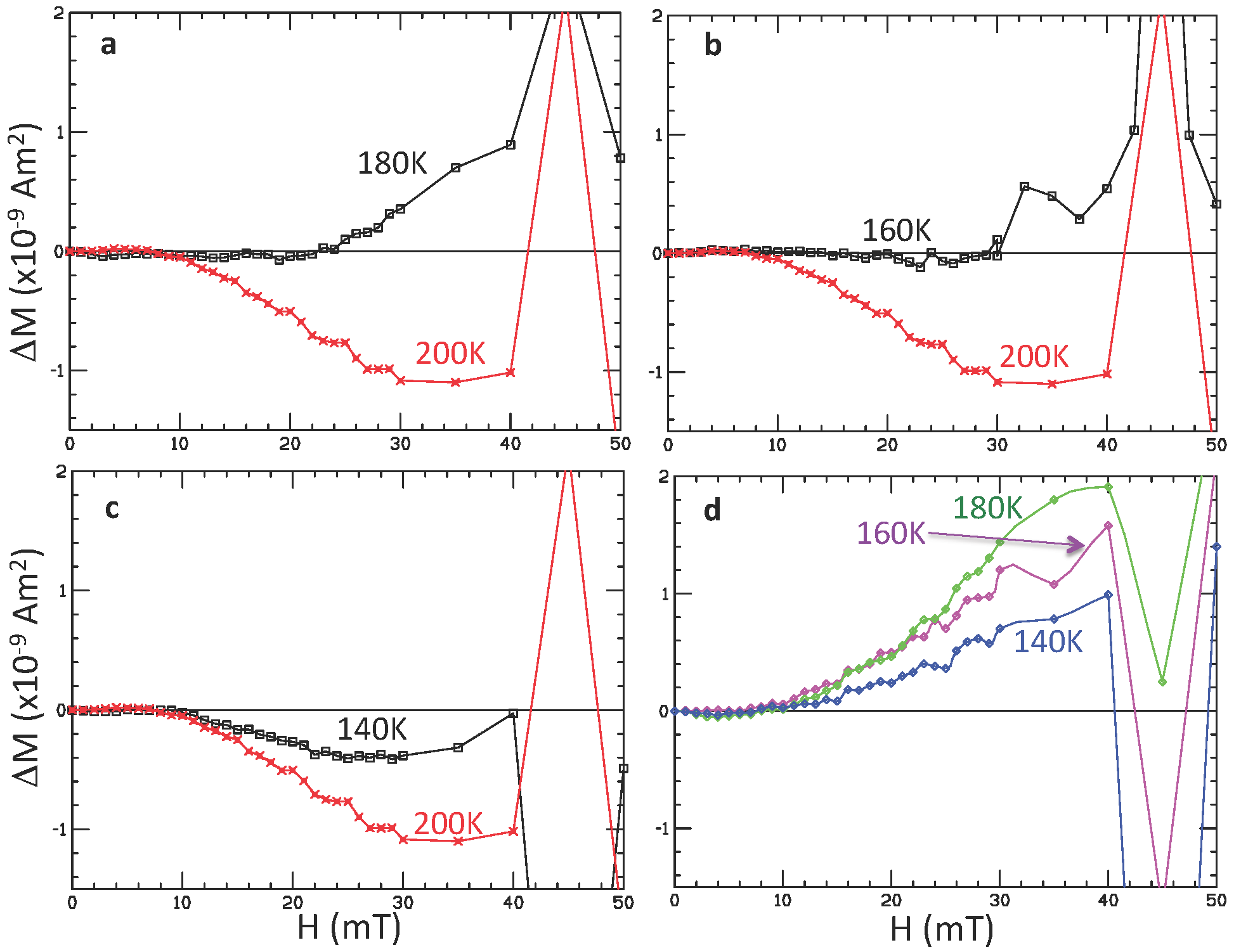}}
		\caption{ {\bf a, b, c}: $\Delta M(H)$ curves for T=180K, 160K, 140K, with offsets
			and ``Meissner slopes'' subtracted. The Meissner slopes, $b=-5.1\times 10^{-7}
			Am^{2}/T$, $b=-5.3\times 10^{-7}Am^{2}/T$ and
			$b=-5.35\times 10^{-7}Am^{2}/T$ respectively, were chosen to fit the linear behavior
			seen in the region $0\le H \le 10mT$. {\bf a, b} and {\bf c} also show
			$\Delta M(H)$ in red, with Meissner slope $b=-4.2\times 10^{-7}$ as used in Ref.
			\cite{reval}. {\bf d} shows the result of subtracting the 200K non-linear background
			from the data for T=180K (green points), 160K (magenta points) and 140K (blue
			points).}
		\label{fig_misfits}
	\end{figure}
	
	In their Supplementary Figure 3a of \cite{reval}, the authors point out that
	the M(H) data for T=200K, above the assumed $T_{c}$ of the $H_{3}S$ sample, the
	signal shows a subtle downturn at approximately 10 mT, and a very similar
	feature is seen for the T=20K data in their Supplementary Figure 3b. This suggest
	that in order to determine the point where the moment originating from the sample
	itself starts to deviate from linearity, this non-linear background needs to
	be subtracted. In Fig. 19 {\bf a, b, c} we show the results for $\Delta M(H)$
	from subtracting a linear background, as was done by the authors of \cite{reval},
	so that $\Delta M$ is approximately zero in the region $0 \le H \le 10mT$, for
	temperatures T=180K, 160K and 140K. The authors inferred that the deviation from
	linearity occurred for $H_{p}=20mT$ for 180K, $H_{p}=30mT$ for 160K and
	$H_{p}>30mT$ for T=140K. Instead, if we subtract the 200K non-linear
	background from the data we find what is shown in Fig. 19 {\bf d}, namely that
	the point where the data start to deviate from linearity for all these
	temperatures is approximately 10mT. This does not support the interpretation that
	deviation from linearity gives the value of the lower critical field of a superconducting
	sample in these experiments, since that value should increase as the temperature decreases.
	
	\subsection{Conclusions on the validity of the revaluation procedure}
	
	The lower critical field is determined from the point where $M(H)$ deviates by
	a certain cut-off from linear. In their revaluation the authors use a fitting
	procedure that is principally reproducible and takes away some of the
	arbitrariness involved in the earlier `by eye' determination. However, we
	believe that the large noise on the $M(H)$ data precludes any objective
	determination of a deviation from linear. The good correspondence between the original
	and re-evaluated values appears to be the result of adjusting the cut-off value
	and the fitting range, sub-optimal fitting, and possibly the choice of datasets.
	Especially the per $M(H)$ curve selection of cut-off value based on noise
	level is most puzzling. It goes against the fixed-value approach of the original
	publication \cite{talan} and results in an obvious bias towards higher $H_{p}$
	values for the noisier low-temperature curves.

	\section{Conclusions}
	
	In this paper we have analyzed the relation between the results published in
	Ref. \cite{e2021p} and the underlying data reported in \cite{data}. We also
	evaluated the consistency of the revaluation procedure reported in Ref. \cite{reval}
	that claims to lead to results consistent with those of Ref. \cite{e2021p}. We
	have found that:

	(1) The data reported in Ref. \cite{e2021p} are incompatible with the
	underlying measured data of \cite{data} in a variety of ways.
	
%
	(2) The revaluation procedure \cite{reval} uses arbitrary criteria for
	choosing the parameters in the fitting procedure, and the large noise of the data precludes any objective 
	determination of where they  deviate from linearity.
	
	(3) As a consequence of (1) and (2),  the revaluation analysis presented in Ref. \cite{reval} does not
		support the conclusions drawn by the authors in Ref. \cite{reval} nor Ref. \cite{e2021p}.
	
	(4) Questions about the consistency of a variety of statements in Refs. \cite{e2021p},
	\cite{correction}, \cite{reval} with reality remain open.
	
	\section{Supplementary information}
	
	\renewcommand{\thefigure}{S\arabic{figure}}
	\setcounter{figure}{0}
	
	\begin{figure*}
		\includegraphics[width=\textwidth, height=\textheight, keepaspectratio]{
			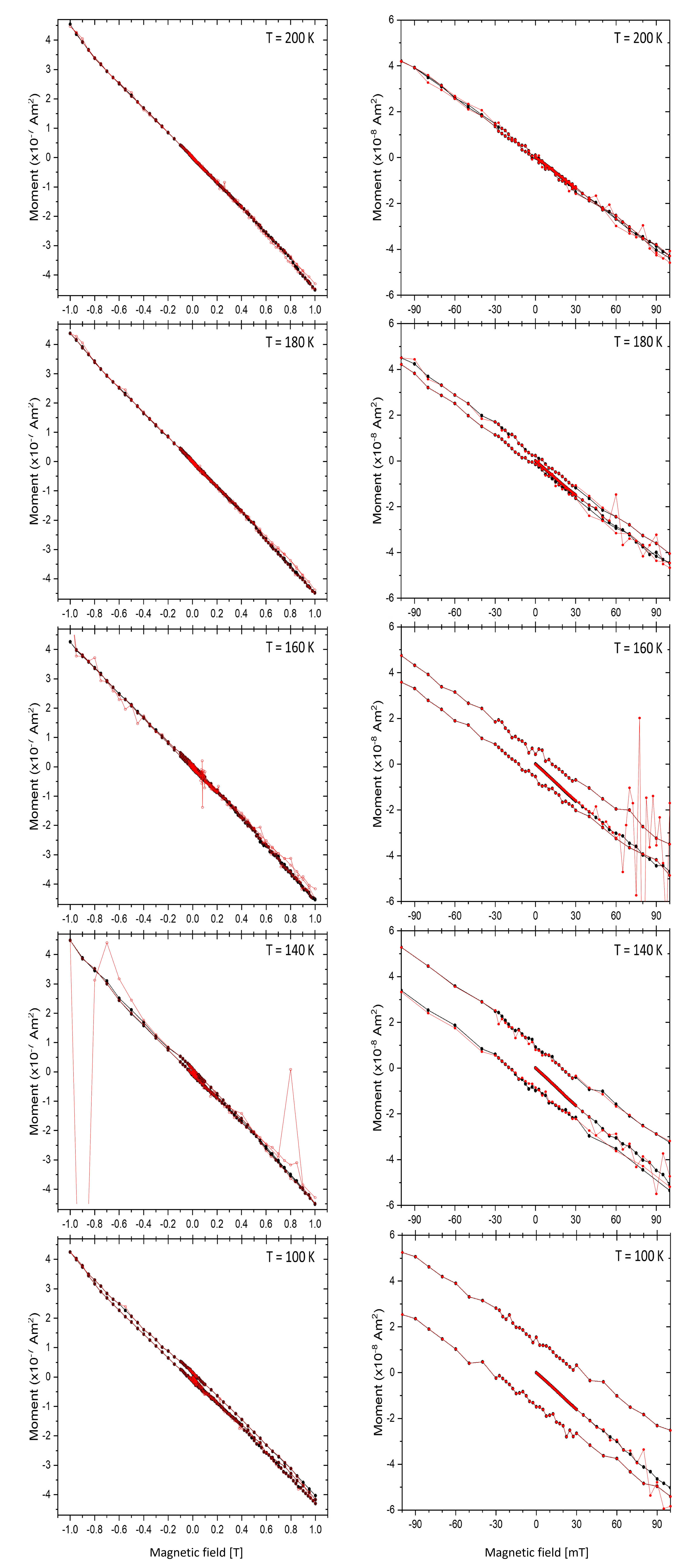
		}
		\caption{ H$_{3}$S magnetization data. Black: Fig. S10 from the Nature
			Communications paper \cite{e2021p}. Red: overlay of the data deposited on
			OSF \cite{data}. }
		\label{fig_s10}
	\end{figure*}
	
	\begin{figure*}
		\includegraphics[width=\textwidth, height=\textheight, keepaspectratio]{
			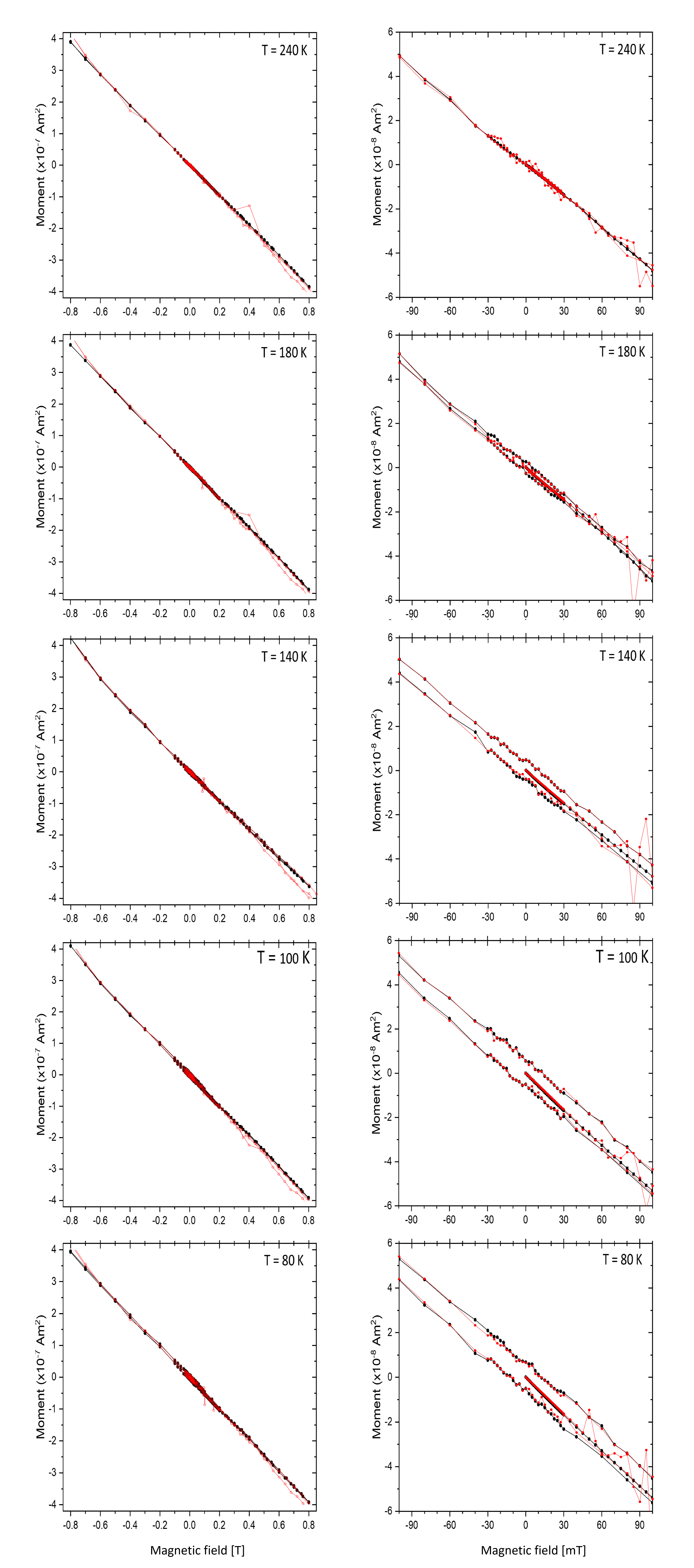
		}
		\caption{ LaH$_{10}$ magnetization data. Black: Fig. S11 from the Nature Communications
			paper \cite{e2021p}. Red: overlay of the data deposited on OSF \cite{data}. }
		\label{fig_s11}
	\end{figure*}
	
	
	\begin{figure*}[]
		\resizebox{11.0cm}{!}{\includegraphics[width=6cm]{
				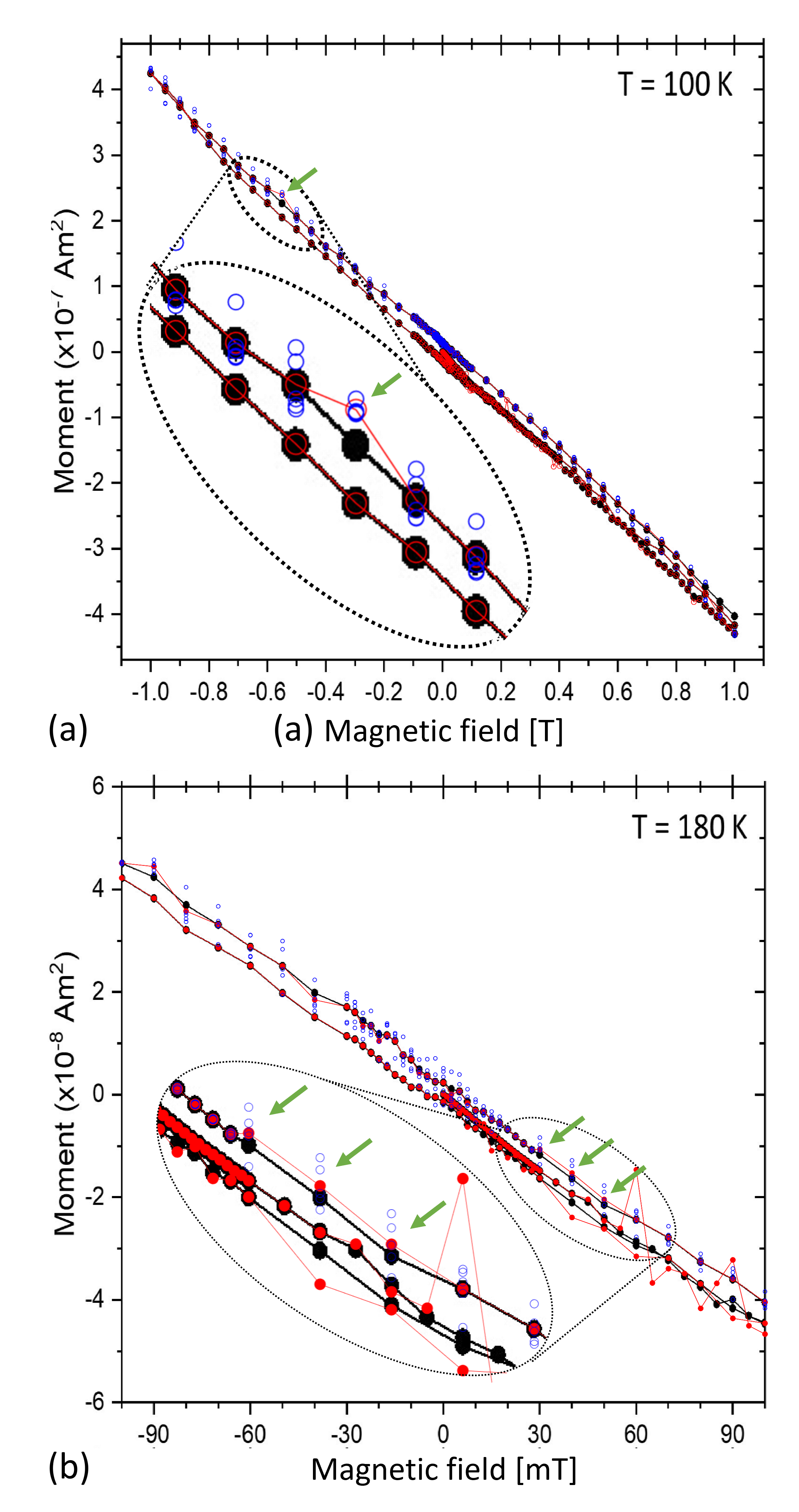
		}}
		\caption{In-depth look at a number of discrepancies. (a) The 100K H$_{3}$S dataset
			from the Nat. Comm. paper in black and of OSF in red. For negative magnetic
			fields only a \emph{single} datapoint at -0.55 T (green arrow) differs
			between the sets. The inset shows the relevant region magnified. Each red
			datapoint is the average of five magnetization measurements that are shown
			by blue symbols. For the -0.55 T point these individual measurements are
			tightly clustered and sit significantly above the black point. Removal of outliers
			can thus not explain the absence of this small hump from Fig. S10 in the Nat.
			Comm. paper. (b) 180K H$_{3}$S OSF and Nat. Comm. magnetization data in a $\pm
			100$ mT range. 28 of the 39 points in the upper hysteresis curve are
			identical, see also Fig. \ref{fig_datasets_h3s_180K}. The green arrows point
			to three sequential datapoints that do differ. The eight points before and
			five after this sequence are all identical. The individual magnetization measurements
			in blue show no outliers that can explain the apparent vertical shift of the
			three Nat. Comm. datapoints.}
		\label{figure1}
	\end{figure*}
	
	\begin{figure*}[]
		\resizebox{18cm}{!}{\includegraphics[width=6cm]{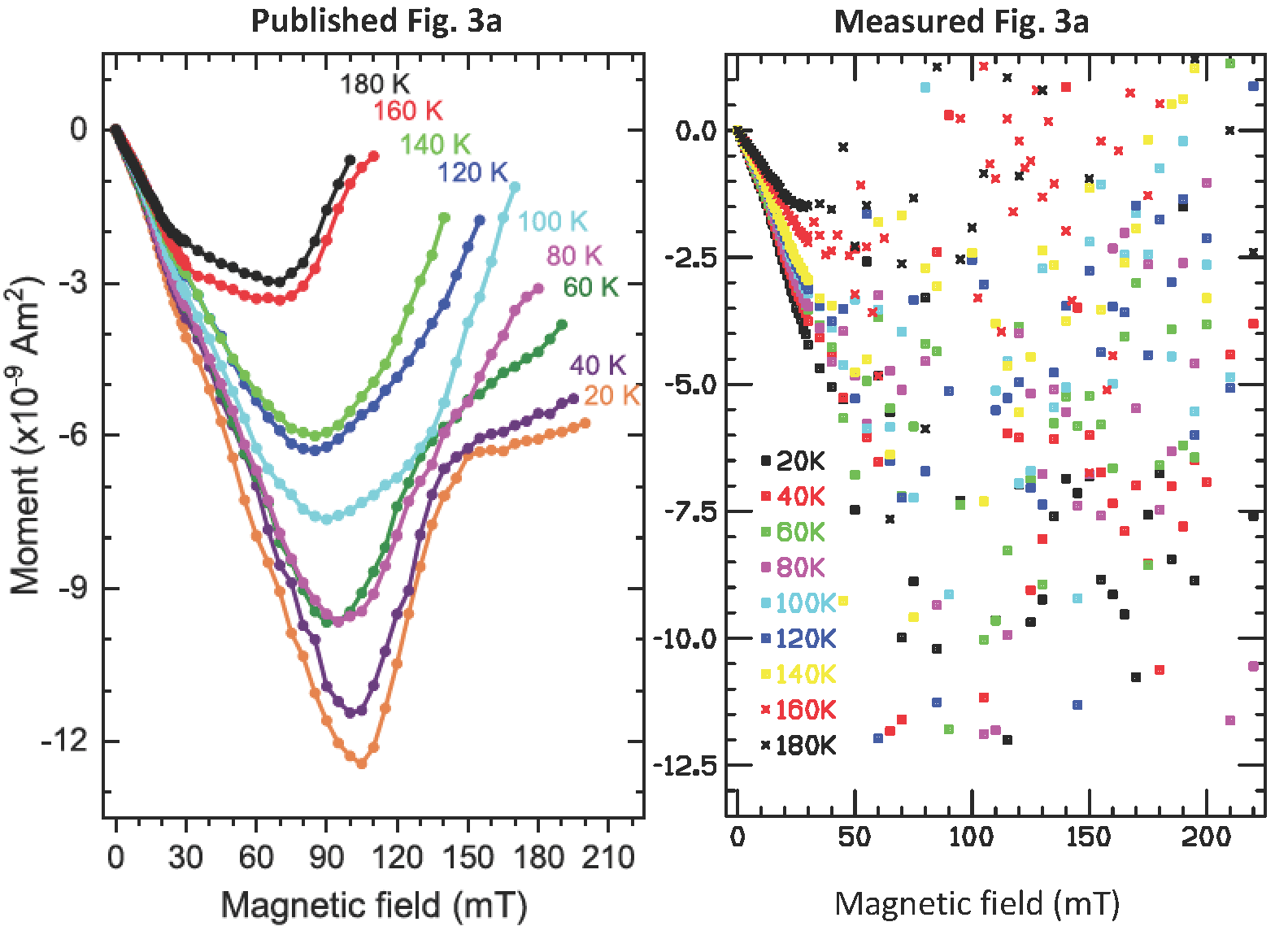}}
		\caption{Data published in Ref. \cite{e2021p} (left panel) and data inferred
			from the measured data after subtraction of a linear background connecting data
			points at magnetic field 0T and 1T, as described in Ref. \cite{correction}.
			We use the same scale on right and left panels to facilitate comparison. How
			the left panel was obtained from the right panel is unknown.}
		\label{figure1}
	\end{figure*}
	
	\begin{figure*}[]
		\resizebox{18cm}{!}{\includegraphics[width=6cm]{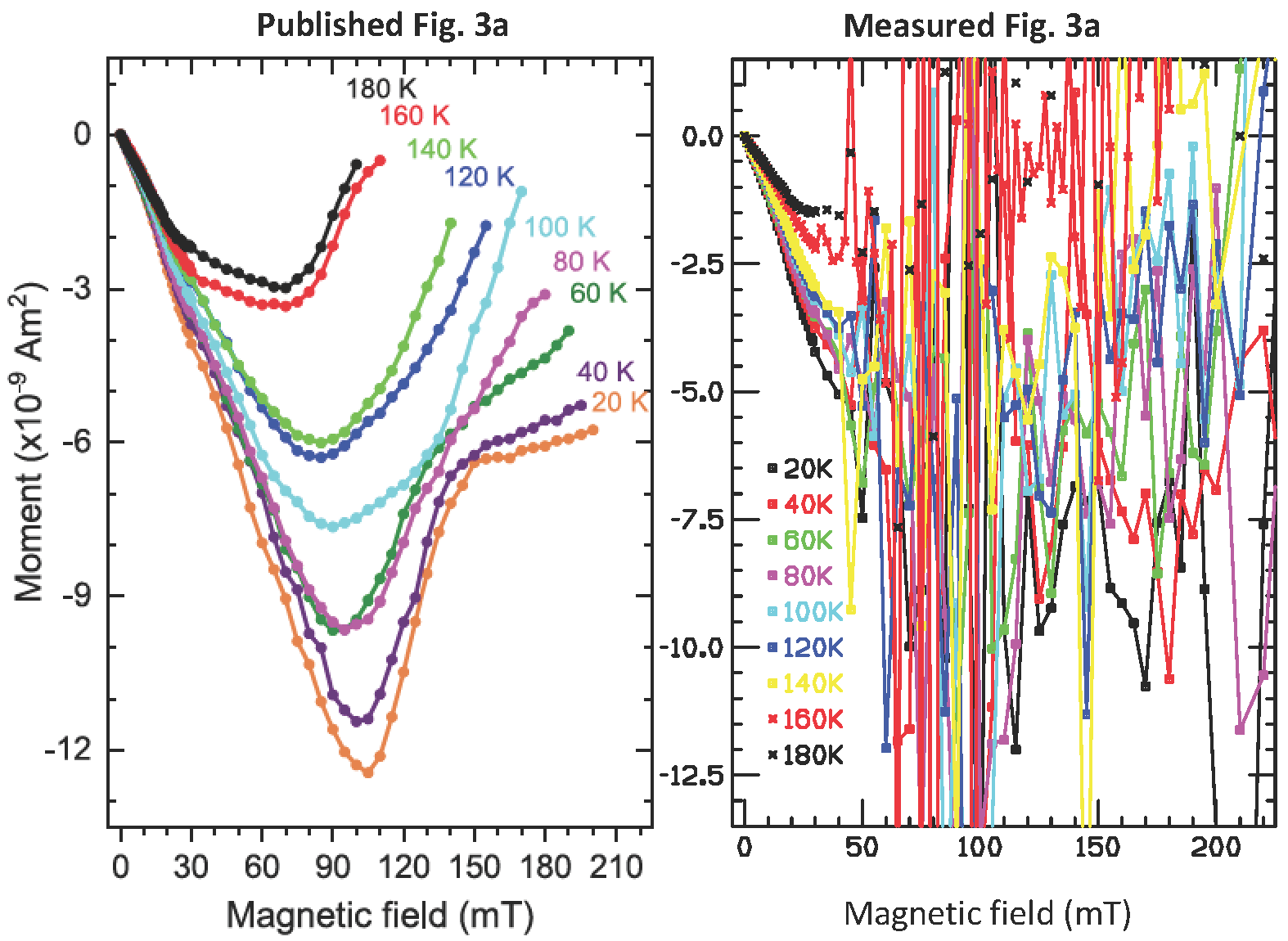}}
		\caption{Same as Fig. S4, except that on the right panel neighboring points
			for each temperature value are connected by straight lines, as done on the left
			panel. How the left panel was obtained from the right panel is unknown.}
		\label{figure1}
	\end{figure*}

\end{document}